   \documentstyle[aps,prl,twocolumn,epsfig]{revtex}

\begin{document}
 
   \draft

   %Fixing abstract in twocolumn mode

   \twocolumn[\hsize\textwidth\columnwidth\hsize\csname 
@twocolumnfalse\endcsname%

   %\twocolumn

   \title{Statistical analysis of air and sea temperature anomalies}
   \author{Nicola Scafetta$^{1,2}$, Tim Imholt$^{2}$, Paolo 
Grigolini$^{2,3,4}$, and Jim Roberts$^{2}$.}
\address{$^{1}$Pratt School EE Dept., Duke University,  P.O. Box 90291, 
Durham, North Carolina 27708 }
   \address{$^{2}$Center for Nonlinear Science, University of North 
Texas,
   P.O. Box 311427, Denton, Texas 76203-1427 }
      \address{$^{3}$Dipartimento di Fisica dell'Universit\`a di Pisa 
and
   INFM, Piazza Torricelli 2, 56127 Pisa, Italy}
   \address{$^{4}$Istituto di Biofisica CNR, Area della Ricerca di 
Pisa,
   Via Alfieri 1, San Cataldo 56010 Ghezzano-Pisa, Italy}
   \date{\today}
   \maketitle

   \begin{abstract}
 This paper presents a global air and sea temperature anomalies 
analysis 
based upon a combination of the wavelet multiresolution analysis and 
the 
scaling analysis methods of a time series.  The wavelet multiresolution 
analysis decomposes the two temperature signals on a {\it 
scale-by-scale} 
basis. The {\it scale-by-scale} smooth and detail curves are compared 
and the 
correlation coefficients between each couple of correspondent sets of 
data 
evaluated. The scaling analysis is based upon the study of the 
spreading and 
the entropy of the  diffusion generated by the temperature signals. 
Therefore, we jointly adopt two distinct methods: the Diffusion Entropy 
Analysis (DEA) and the Standard Deviation Analysis (SDA). The joint use 
of 
these two methods allows us to establish with more 
confidence
the nature of the signals, as well as their scaling, and it yields 
the discovery of a slight L\'{e}vy component in the two temperature 
data sets.
 Finally, the DEA and SDA are used to study the wavelet residuals of 
the two 
temperature anomalies. The temporal regions of persistence and 
antipersistence  of the signals are determined and the non-stationary 
effect 
of the 10-11 year solar cycle upon the temperature is studied.  The 
temperature monthly data  cover  the period from 1860 to 2000 A.D.E.  

 \end{abstract}
   \vspace{0.5cm}
   %
   %Fixing abstract in twocolumn mode%%%
   ] 
   %%%%
   %

\section{Introduction}

The statistical analysis of time series is a challenging problem of 
statistical mechanics. This is due to the fact that there are still 
many 
unsettled problems. The most important seems to be that the techniques 
of 
analysis that are currently used are based on the assumption that the 
time 
series under study are generated by stationary processes. In general 
this is 
not the case.  The time series mirroring complex processes are usually 
non-stationary in nature. The non-stationary condition seems to be a 
very 
general property, although
it has any number of possible sources in any system. For instance the 
origin 
of non-stationarity in the case of solar flares is given by the solar 
cycles 
(for a recent review about this interesting issue, see Ref. 
\cite{solarcycle}) and a special caution must be adopted to take the 
effects 
of this non-stationarity into account\cite{solarflares}. In fact, it 
has been 
recently shown \cite{thermodynamicofsocialprocesses}  that the memory 
left 
after detrending
annual periodicity is much less intense than imagined in earlier 
publications\cite{morememory}. 
Another issue, which seems to be still unsettled, is as to the 
statistical 
nature of the fluctuations, once their genuinely stationary nature has 
been 
assessed. Are these fluctuations Gaussian? Are these fluctuation of 
L\'{e}vy 
in nature?

In this paper we want to illustrate an efficient approach to the 
solution of 
these difficulties. To stress the efficiency of this approach we apply 
it to 
the analysis of global air and sea temperature anomalies, a problem 
where, as 
we shall see, properly detrending non-stationary components is 
an essential request to shed light into the nature of the process under 
study.  
The approach we intend to use rests on the joint use of the Diffusion 
Entropy 
Analysis (DEA) and wavelet analysis of time series.  DEA was born
as an efficient way to detect 
scaling\cite{thermodynamicofsocialprocesses,dea2,scalingdetection}, 
with 
applications to sociological\cite{thermodynamicofsocialprocesses} and  
astrophysical\cite{solarflares} processes.  This technique of analysis 
has 
been applied with success also to the study of DNA 
sequences\cite{dea3,dea4} 
and  heart beat rhythms in cardiac patients \cite{dea5}. Furthermore, 
some 
attention has been devoted to 
establish the connection between DEA and the Kolmogorov 
complexity\cite{dea6} 
and it is becoming clear that this technique can also be used to study 
the 
transition from dynamics to thermodynamics, a crucial property that is 
used 
with success to study small portions of large sequences\cite{dea3}, 
thereby 
establishing a possible way to address the problem of non-stationality.  
Research work is currently being done to make it possible to utilize 
this 
technique to address the cases of multiple scaling\cite{dea7}.   

 Wavelet techniques are a powerful method of 
analysis\cite{percival} that localizes a signal simultaneously in time 
and 
frequency.  We use wavelets  for the purpose to decompose the signal in smooth, detail and residual components.
The wavelet decomposition has been shown to be an efficient way of
detrending from the 
data a 
non-stationary component in a natural way, so as to bypass the main 
difficulties concerning the non-stationary nature of the data under 
study \cite{scaffbruc}. 
The adoption of DEA makes the  scaling emerge and also sheds light into 
the 
statistical nature of the fluctuations around the non-stationary bias.

Let us now illustrate the time series under study in this paper.  The 
time 
series of annually averaged global surface temperature anomalies have
attracted the attention of many scientists since the pioneering work of 
Nicolis and Nicolis\cite{nicolis}. Dynamical systems theory has 
provided a 
new quantitative perspective on the predictability of weather and 
climate 
processes. For more recent attempts along these lines the interested 
reader 
can consult the work of Ref.\cite{gimeno}. The conceptual paradigm 
behind our 
analysis is that of intermittence as a dynamic source of L\'{e}vy 
statistics\cite{dea7}. The detection of L\'{e}vy 
scaling\cite{scalingdetection} after detrending the non-stationary 
component 
would corroborate the validity of an intermittent dynamic model.

 It is worth remarking that the term \emph{temperature anomalies} is a 
technical definition
adopted in the current literature on weather and climate processes on 
earth 
to denote air and sea temperature departures from a mean temperature 
value. 
Therefore, this term must not be confused with the term anomalous 
diffusion 
that is related to our conceptual paradigm. We use the data on 
temperature 
fluctuations to generate a diffusion process that is compared to the 
standard 
Brownian motion. The departure of this resulting diffusion process from 
Brownian diffusion is
called anomalous diffusion. The DEA aims at measuring the strength of 
this 
anomaly. Consequently, we can say that one of the aims of this paper is 
to 
determine the anomalous nature of the diffusion process generated by 
air and 
sea temperature anomalies.
The data analyzed are updated continuously by the Climate Research Unit 
in 
the United Kingdom\cite{cru} and the Hadley Centre for Climate 
Prediction and 
Research, Meteorological Office\cite{hcc}.  The earliest of attempts at 
collecting these data were done in 1986\cite{global} and has culminated 
in 
what is recognized as one of the most accurate data files for global 
air 
temperature, and global sea surface temperature (SST)\cite{jones}. The 
land-air temperature has been corrected for non-climatic errors, such 
as 
changes in the location of the weather stations and changes in instrumentation\cite{jones1}.  
The SST data have been corrected for changes in 
instrumentation that was used before 1942 
\cite{parker,parker1,folland}. The 
data consists of numbers that represent the departure from a mean 
temperature 
in order to see a change from a global average of sorts.  The mean 
temperature used was the 1961-1990 mean temperature.

Figs. 1 show the global air (a) and sea (b) temperature anomalies in 
the 
period from 1860 to 2000 A.D.E. The dashed lines are the wavelet 
multiresolution S7 smooth portion of the signal.  The exact 
mathematical 
definition of wavelet smooth curve is given in Section 3. To understand 
the 
meaning of the data illustrated in Fig. 1, it is enough for the reader 
to 
consider the S7 smooth as a convenient way to establish a type of mean 
value 
about which the temperature fluctuations take place. In fact, as it 
will 
become clear in Section 3, the S7 smooth is obtained by a wavelet 
average of 
the data over time intervals of 128 months. 

The global air and sea temperature anomalies are very similar to one 
another. 
The mean value of temperature fluctuation shows that from 1860 to 1915 
the 
average temperature is almost constant, with a change of only $0.2^o$ Celsius. 
From 1915 to 1950 the temperature increases by $0.6^o$ Celsius. From 1945 
to 1980 
the average temperature remains almost constant again.  Finally, from 
1980 to 
current time there is a further increase of the average temperature of 
almost 
$0.4^o$ Celsius. What about the fluctuations around this mean value? Figs. 
2 show 
the spectral density against the period in months of the global air (a) 
and 
sea (b) temperature anomalies in the period from 1860 to 2000 A.D.E.. 
Some of 
these periodicities are reported in Table I. The main periodicities 
involve a 
time period  6 and 12 months long (related to the yearly cycle of the 
Earth 
orbiting the Sun), 9-12 years long, 21-22 years long and, finally,  a 
strong 
periodicity 55-57 years long (the last three cycles are all established 
solar 
cycles with the 9-12 year cycle being the most widely known and 
observed 
one). 

This paper aims at settling several questions concerning these data. 
The 
correlation among air and sea temperature smooth curves seems to be 
evident. 
However, it is not so clear if these correlations exist also at the 
level of 
fluctuations. Is there anomalous scaling? If there is anomalous 
scaling, does 
it rest on smooth curves or fluctuations? Which is the statistical 
nature of 
these fluctuations? Are they Gaussian fluctuations? Are they L\'{e}vy 
fluctuations?

The outline of the paper is as follows. In Section 2, to make the paper 
as 
self contained as possible, we make a short review of wavelet 
multiresolution 
analysis. Section 3 is devoted to the study of the correlation between 
global 
air and sea temperature anomalies at the scale of both the wavelet 
smooth and 
detail curves. Section 4 is a short review of the DEA and shows this 
technique at work on the data without any decomposition. Section 5 
illustrates the joint use of wavelets and DEA.  Finally, in Section 6 
we make 
a balance on the results obtained in this paper.

\section{Wavelet multiresolution analysis}

Wavelet analysis  \cite{percival} is a  powerful method to analyze time 
series that is attracting the attention of an ever increasing number of 
investigators. 
Wavelet Transform makes use of scaling functions, the wavelets,   which 
are 
characterized by the important 
property of being localized in both time and frequency. These 
functions integrate to zero and, usually, are normalized.
A scaling
coefficient $\tau$ characterizes a wavelet. The length $2\tau$ measures 
the 
width of the wavelet and defines the time scale analyzed by the 
wavelet. Two 
typical
wavelet functions that are widely used in  the continuous 
wavelet 
transform are the Haar wavelet 
and the Mexican hat wavelet \cite{percival}.
The Haar wavelet is defined as: 
\begin{equation}\label{haarwds}
^{\left(H\right)}\tilde{\psi}_{\tau,t}(u)\equiv \left\{
\begin{array}{ccc}
-1/\sqrt{2\tau},          & ~~t-\tau<u<t    \\
1/\sqrt{2\tau},          & ~~t<u<t+\tau   &    \\
 0,         & ~~otherwise   &.  \\
\end{array}\right.
\end{equation}
The Mexican hat wavelet is the second derivative of a Gaussian 
function.
 Given a signal $\xi(u)$, the Continuous Wavelet Transform  is defined 
by
\begin{equation}\label{cwtdhjj3}
W(\tau, t)=\int\limits_{-\infty }^{\infty } \tilde{\psi}_{\tau 
,t}(u)~\xi(u)~du~.
\end{equation}
The original signal can be recovered from its Continuous Wavelet 
Transform via
\begin{equation}\label{invcwtxu9}
\xi(u)=\frac{1}{C_{\tilde{\psi}}} \int\limits_{0}^{\infty} \left[ 
\int\limits_{-\infty }^{\infty } W(\tau,t) ~\tilde{\psi}_{\tau,t}(u)~dt 
\right] ~\frac{d\tau}{\tau^2}~,
\end{equation}
where $C_{\tilde{\psi}}$ is a constant that depends on the wavalet 
function 
\cite{percival}.
The double integral of Eq. (\ref{invcwtxu9}) suggests that the original 
signal may be decomposed in ``continuous details'' that depend on the 
scale 
coefficient $\tau$. However, there exists a discrete version of the wavelet transform, the Maximum 
Overlap 
Discrete Wavelet Transform (MODWT), which is the basic tool needed for 
studying time series of $N$ data via wavelet. In the Ref.  
\cite{percival}, 
the reader can find all of the mathematical details.      For the 
purpose of 
this paper, it is important to have in mind only one of the important 
properties of the MODWT: the  Wavelet Multiresolution Analysis (WMA). 
It is 
possible to prove that given an integer $J_0$ such that $2^{J_0}<N$, 
where 
$N$ is the number of the data points, the original time series 
represented by 
the vector ${\bf X}$ can be  decomposed as follows:
\begin{equation}\label{decomw}
{\bf X}=S_{J_0} + \sum _{j=1}^{J_0} D_j~, 
\end{equation}
with 
\begin{equation}\label{decomrel}
S_{j-1}= S_{j} + D_j~.
\end{equation}
The detail $D_j$ represents changes on a scale of $2\tau=2^{j}$, while 
the 
smooth $S_{J_0}$ represents averages on  a scale of $2\tau_{J_0}= 
2^{J_0}$.  
We term wavelet \emph{residuals}  the quantities
\begin{equation}\label{decomwR}
R_{J_0}={\bf X}-S_{J_0} = \sum _{j=1}^{J_0} D_j~.
\end{equation}
It is then evident that we can interpret the residuals as fluctuations 
about 
the local mean value evaluated on the time scale $2\tau_{J_0}= 2^{J_0}$. At 
this 
stage, the reader should fully understand the comments made in Section 
1 
about the data illustrated in Fig. 1. 

For the reader to properly appreciate the value of WMA, we show this 
technique at work by means of the results illustrated in Figs. 3 and 4. 
These 
figures  show the WMA of the global air and sea temperature anomalies   
in 
the years 1860-2000. The analysis is done by using the Daubechies {\it 
least 
asymmetric}  scaling wavelet filter (LA8) \cite{percival}. The LA8 
wavelets 
look similar to the Mexican hat but they are asymmetric, a fact that 
makes 
them more plastic than the Mexican hat wavelet.   We have plotted the 
WMA for 
$J_0=7$.  Figs. 3 compare the smooth curves S4, S5, S6 and S7 and the 
detail 
curves D4, D5, D6, D7 of the air (solid lines) and sea (dashed lines) 
temperature data. Figs. 4  show the details D1, D2, D3 and D4 of the 
two sets 
of data.   Details $D_j$  show the fluctuations of the temperature on a 
scale 
of $2\tau=2^j$ months. According to Eqs. (\ref{decomw}) and 
(\ref{decomrel}), 
the sum of all seven details and the $S7$ smooth give the two original 
signals. The $S6$ smooth curve is given by $S7+D7$; the $S5$ smooth  is 
given by 
$S6 + D6$, and so on, till to $S1$, given by $S2 + D2$. 

Figs. 3 show that the global air and sea temperature anomalies are 
closely 
correlated to each other but they do not coincide. If it happens that 
in a 
given time region air is hotter than sea, immediately afterward the 
opposite 
effect takes place, and sea is warmer than air. The smooth curves of 
Figs. 3, 
and especially the  S7 and S6 smooth curves, show that in the time 
regions  
of  persistent temperature increase (time regions 1860-1880, 1910-1950 
and 
1980-2000) air is hotter. Instead, in the  time regions  of  persistent 
temperature decrease  (time regions 1880-1910 and 1950-1980) air is 
colder. 
This may be explained by the fact that heat capacity of water  is 
higher than 
heat capacity of air, thereby implying that it takes more time for 
water 
temperature either to increase or to decrease.  The D4, D5, D6, D7 
details 
illustrate the  temperature fluctuations  of both air and sea 
corresponding 
to  the time scales of 16, 32, 64 and 128 months respectively.  The  
fluctuations of both kinds of data, air and sea, look remarkably 
similar. The 
analysis of the details D1, D2 and D3 of Figs. 4 show, instead, that 
the 
fluctuations of the air temperature are larger than those of the sea 
temperature. This effect, too, may be related to the higher heat 
capacity of  
water. Finally,   Figs. 4 show a stronger fluctuation of the data 
during the 
period 1860-1880. There is also a strong fluctuation of  sea 
temperature in 
the D3 detail, in the  1920-1950 time region.   Further study is 
required to 
assess whether these stronger fluctuations are due to natural phenomena 
or to 
some artifact of  a non conventional way of data acquisition.

\section{Multiresolution correlation analysis}
The use of the Multiresolution Correlation Analysis via wavelet is a 
simple 
procedure \cite{tim1}. We decompose the two temperature datasets into 
rests 
and details by using 
WMA as demonstrated  in the previous section. Then, we create pairs of 
partners, the first component of the pair being either a rest or a 
detail of 
the sea temperature data and the second  the
corresponding rest or detail of the air temperature data. For any given 
pair 
of  datasets $(x_i, y_i)$; $i= 1, ..., N$, the linear correlation
coefficient $r$ is given by the formula
\begin{equation}\label{corrcoeff}
r=\frac{\sum _{i} \left(x_i-\overline{x} \right)\left(y_i-\overline{y} 
\right)}
{\sqrt{\sum _{i} \left(x_i-\overline{x} \right)^2}\sqrt{\sum _{i} 
\left(y_i-\overline{y} \right)^2}}~,
\end{equation}
where, as usual, $\overline{x}$ is the mean of the former sequence and, 
$\overline{y}$ is the mean of the latter sequence.
The value of $r$ lies between  -1 and 1, and can include the extreme 
values 
$1$ and $-1$. It takes on a value of 1, termed
``completely positive correlation,'' when the data points lie on a 
perfect 
straight line
with positive slope, with x and y increasing together. The value 1 
holds 
independently
of the magnitude of the slope. If the data points lie on a perfect 
straight 
line with
negative slope, y decreasing as x increases, then r has the value  -1; 
this 
is called
``completely negative correlation.'' A value of $r$ near zero indicates 
that 
the variables
x and y are uncorrelated. The two sequences are considered to be 
significantly correlated when the value of $|r|$ is close to $1$. 

 Fig. 5 illustrates the correlation coefficient, $r$, between air and 
sea 
temperature anomalies as a function of the wavelet index $j$, and thus, 
for 
the reasons illustrated in Section 3, as a function of the wavelet 
scale 
$2\tau = 2^{j}$.  The top curve denotes the correlation between  the  
air and 
the sea $S_{j}$ smooth curves, with $j$ ranging from $0$ to $10$. The 
bottom 
curve denotes the coefficient of correlation between air and sea 
details, 
$D_{j}$, with $j$ ranging from $1$ to $10$. The  value $r=0.87$ 
corresponding 
to the top curve at $j = 0$  is the correlation coefficient between the 
two 
original temperature data without any filtering. For the sake of 
reader's 
convenience the values of $r$ are reported   in Table II. From Fig. 5 
we see 
the small details D1, D2, D3 and D4 are not significantly correlated. 
These 
details  refer to a time scale until 16 months, and consequently we can 
conclude that within this time range there are no significant 
correlations 
between sea and temperature fluctuations. The correlation between the 
details 
increases with increasing  the scale index and becomes significant for 
D5, D6 
and D7. The reader can consult   Figs. 3, which confirms in fact 
visually 
that these details are significantly correlated.  The correlation 
between the 
 D7 details yields  a local maximum, $r=0.95$. All this means that the 
S4-S7 
smooth curves are the best indicators of the correlations between  air 
and 
sea temperature anomalies. In fact, according to the prescriptions of 
Section 
3, the S4-S7 smooth curves are obtained by adding to S4 the D5, D6 and 
D7 
details, respectively, and these details, as shown by the bottom curve 
of 
Fig. 5 are correlated. Again, the reader can make a visual inspection 
of 
these smooth curves by consulting Figs 3. Details smaller than the D5 
detail 
have small correlations. This also means that to shed light on the 
properties 
that make sea temperatures different from air temperatures we must 
focus on 
small details.  Smooth curves larger than  S7 are very well correlated 
but do 
not afford additional  information, because they are too smooth.

\section{Scaling analysis}
Scale invariance has been found to hold empirically for a number of 
complex 
systems \cite{2Mandelbrot} and the correct evaluation of the scaling 
exponents is of fundamental importance to assess if universality 
classes 
exist \cite{stanley}.
A widely used method of analysis of complexity rests on the assessment 
of the 
scaling exponent of the diffusion process generated by a time series.  
See, 
for instance, Refs. 
\cite{thermodynamicofsocialprocesses,dea2,dea3,dea4,dfa}.
According to the prescription of Ref. \cite{dfa}, we interpret the 
numbers of 
a time series as generating diffusion fluctuations and we shift our 
attention 
from the time series to the probability distribution function (pdf) 
$p(x,t)$, 
where $x$ denotes the  variable collecting the fluctuations. The 
scaling 
property takes on the form
\begin{equation}\label{scafun12}
p(x,t) = \frac{1}{t^{\delta}}~F\left( \frac{x}{t^{\delta}}\right)~,
\end{equation}
where $\delta$ is the scaling exponent.

\subsection{Gauss and L\'{e}vy diffusion}

There are two main forms of anomalous diffusion. The first is the 
generalization of Brownian motion, proposed years ago by 
Mandelbrot\cite{2Mandelbrot}, known as Fractional Brownian Motion (FBM) 
and 
yielding for the diffusion process a variance increasing in time as 
$t^{2H}$. 
This kind of anomalous diffusion
fits the scaling definition of Eq.(\ref{scafun12}) with $\delta = H$ 
and 
$F(y)$ being a Gaussian function of $y$. A second form of anomalous
diffusion is obtained by generalizing the Central Limit Theorem (CLT). 
The 
prescriptions of  the Generalized Central Limit Theorem (GCLT) 
\cite{gnedenko} are as follows. 
Let us assume that the diffusing variable $x$ is the sum of $t$ 
independent 
random variables $\xi_{i}$,  each of which has a probability 
distribution, 
symmetric around $\xi = 0$. Let us assume also that for large values of 
$|\xi|$, this distribution is an inverse power law, with index $\mu>1$, 
so as 
to fit the normalization condition, and $\mu <2$, so as to violate the 
CLT 
constraint. Then, according to the GCLT, for $t \rightarrow \infty$, 
the 
diffusion becomes stable and the Fourier transform of $p(x,t)$ gets the 
form
\begin{equation}\label{carlevyflu}
\hat{p}(k,t)=\exp \left(- b |k|^{\alpha }t\right),
\end{equation}
where $b$ is a kind of generalized diffusion coefficient, determined by 
the 
strength of the fluctuations and $\alpha = \mu +1$. It has been shown 
\cite{gnedenko} that there is scaling. This means that the resulting 
diffusion process fits the condition of Eq. (\ref{scafun12}),
with $\delta = 1/\alpha$, namely,
\begin{equation}\label{scalrelD}
\mu=1+\frac{1}{\delta}.
\end{equation}

Both forms of anomalous diffusion are idealization of reality. The 
theory of  
FBM implies that the corresponding scaling property holds true at any 
time 
scale, while its dynamic derivation\cite{floriani}, in the persistent 
case $H 
> 0.5$,  suggests that it is a time asymptotic property generated by 
the 
fluctuations of a Gaussian variable $\xi(t)$, whose correlation 
function has 
inverse power law with index $\beta < 1$ ($H = 1-\beta/2$). 
The main problem with this dynamic interpretation is that there are no 
stable 
theorems behind the Gaussian property of the ``microscopic" fluctuation 
$\xi(t)$. It is necessary to supplement the dynamic model with random 
ingredients that do not have a dynamic origin\cite{buiatti}.
The earlier illustrated approach to L\'{e}vy statistics, called  
L\'{e}vy  
flight, is judged to be unrealistic, because it involves a random 
velocity
that can be arbitrarily large. To bypass this difficulty recourse is 
given to 
L\'{e}vy walk\cite{zumofenklaftershelsinger}. We adopt the following 
dynamic 
model\cite{dea7}. Let us consider a sequence $\{\tau_{i},s_{j}\}$, with 
$i 
=0, 1,...,\infty$. The numbers $\tau_{i}$ are random numbers with the 
distribution density
\begin{equation}\label{densitydistributionoftau}
\psi(\tau) = (\mu-1) \frac{T^{\mu-1}}{(T  + \tau)^{\mu}}~,
\end{equation}
where $T$ is a positive constant. Note that to ensure the stationary 
condition the additional condition $\mu >2$ is necessary, and we make 
this 
assumption also in this paper.
Thus, we have $\mu < 3$ to ensure the anomalous character of the 
resulting 
diffusion process and $\mu >2$ to make our dynamic picture stationary. 
It is 
easy to prove that $<\tau> = T/(\mu-2)$. The numbers $s_{i}$ have the 
values 
$1$ and $-1$, determined by the coin tossing rule. 
For a generic time $t$, let us consider the time $t_{N}$ fitting the 
conditions $t_{N} = \tau_{0}+\tau_{1}+...\tau_{N-1} + \tau_{N} < t$ and 
$t_{N} + \tau_{N} > t$. Then the trajectory prescribed by this dynamic 
model 
is $x(t) =  W[ \tau_{0} s_{0} + \tau_{1} s_{1} + ...\tau_{N-1} s_{N-1} 
+(t-t_{N})s_{N}]$.  We see that in this case, due to the large memory 
time, 
the random walkers can travel ahead or backwards by quantities with the 
same 
distribution as the L\'{e}vy flight. However, this takes a time 
proportional 
to the traveled length. The correlation functions of each elementary 
jumps, 
either $W$ or $-W$ has an inverse power law with index $\beta$, which 
is now 
given by $\beta = \mu -2$.  It is important to notice that the process 
is now 
multiscaling \cite{dea7}, due to the fact that the propagation fronts 
propagate linearly in time. The central part of the distribution, if we 
neglect the truncation produced by the finite velocity of the 
propagation 
front, is given by $\delta = 1/(\mu-1)$, namely, by 
Eq.(\ref{scalrelD}).

It is important to stress that this dynamic model to L\'{e}vy 
statistics 
yields, at any finite time $t$, finite second moments. Consequently, it 
could 
be interpreted as a form of FBM. However, in this case $H$ would not 
correspond to the correct scaling of the central part of the 
distribution and 
$\delta$ and $\mu$ are related to this pseudo-scaling $H$ 
by\cite{scalingdetection}.
\begin{equation}\label{relHdelta34}
\delta =\frac{1}{3-2H}~.
\end{equation}
and
\begin{equation}\label{scalrelH}
\mu=4-2H,
\end{equation}
respectively.

\subsection{The diffusion algorithm}

Let us consider a sequence of $N$ numbers
\begin{equation}
    \xi_{i} ,    \quad i = 1,  \ldots ,N.
    \label{thesequenceundestudy}
    \end{equation}
    The goal is to establish the possible 
    existence of a scaling, either normal or anomalous, in the most 
    efficient way as possible without altering the data with any form 
    of detrending. First of all, let us select  an integer number
    $t$, fitting the condition $1 \leq t < N$. 
This integer number will be referred to by us as ``diffusion time''. 
For any 
given 
    time $t$ we can find $M(t)=N - t +1$ sub-sequences defined by
    \begin{equation}
        \xi_{i}^{(s)} \equiv \xi_{i + s}, \quad with  \quad s = 0,  
\ldots ,  
N-t.
        \label{multiplicationofsequence}
        \end{equation}
        For any of these sub-sequences we build up a diffusion 
trajectory, 
$s$, defined by the position        
        \begin{equation}
    x^{(s)}(t) = \sum_{i = 1}^{t} \xi_{i}^{(s)} 
    = \sum_{i = 1}^{t} \xi_{i+s}.   
        \label{positions}
        \end{equation}

The direct evaluation of variance is probably the most natural method 
of 
variance detection. All
trajectories start from the origin $x(t=0)=0$.  With increasing time 
$t$, the 
sub-sequences generate a diffusion
process. At each time $t$, it is possible to calculate  the
Standard Deviation of the position of the $M(t)$  sub-sequences with 
the 
well known expression:
\begin{equation}\label{varvar33}
D(t) =\sqrt{\frac{ \sum _{s=0}^{N-l}\left[x^{(s)}(t)-\overline{x}(t) 
\right]^2}{M(t)-1}},
\end{equation}
where $\overline{x}(t)$ is the average of the positions of the
$M(t)$ sub-trajectories at time $t$. The exponent $H$ is defined by 
\begin{equation}\label{scaliH}
D(t)\propto t^H ~.
\end{equation}
We call this approach to  the scaling evaluation Standard Deviation 
Analysis 
(SDA). In Ref.\cite{scalingdetection} the interested reader can find an 
illustration of the traditional techniques of scaling detection and of 
why 
all of them are virtually equivalent to the SDA. 

The DEA, is based upon the following algorithm. We have to partition 
the 
$x$-axis into cells of size 
       $\epsilon(t)$. When this partition is made, we have to label the 
       cells. We count how many particles are found in the same cell at 
a 
       given time $t$. We denote this number by $N_{i}(t)$. Then 
       we use this number to determine the probability that a particle 
       can be found in the $i$-th cell at time $t$, $p_{i}(t)$, by 
means 
       of
       \begin{equation}
        p_{i}(t) \equiv  \frac{N_{i}(t)}{M(t)} .
        \label{probability}
        \end{equation}
        At this stage the entropy of the diffusion process at the time 
$t$
        is determined and reads
\begin{equation}
 S(t) = - \sum_{i} p_{i}(t) ~\ln [p_{i}(t)]~.
\label{entropy}
        \end{equation}
        The easiest way to proceed with the 
choice of the cell size, $\epsilon(t)$, is to assume it to be a 
fraction of 
the square root of 
the variance of the 
fluctuations $\xi(i)$, and consequently independent of $t$. 
If the scaling condition of Eq. (\ref{scafun12}) holds true, it is easy 
to 
prove that 
\begin{equation}\label{scafun14}
S(t) = A + \delta~  \ln (t) ~,
\end{equation}
        where, 
in the continuous approximation,
        \begin{equation}
        A \equiv -\int_{-\infty}^{\infty} dy \, F(y) \, \ln [F(y)]~,        
\label{ainthecontinuouscase}
        \end{equation}
with  $y = x/t^{\delta}$.
The scaling Eqs. (\ref{scaliH}) and  (\ref{scafun14}) determine the 
exponents 
$H$ and $\delta$.

\subsection{Data analysis}

 Fig. 6 shows the numerical results by using the DEA (Figs. 6a and 6c) 
and 
the SDA (Figs. 6b and 6d) of the global air (Figs. 6a and 6b) and sea 
(Figs. 
6c and 6d)  temperature anomalies. In the ordinate axis we plot 
$D(t)/D(1)$ 
and $S(t)-S(1)$. Thus,  the curves start from 1 and 0, respectively. 
The 
straight lines are function of the type $f_{DE}(t)=\delta ~\ln(t)$ and 
$f_{SD}(t)=t^H$ and become straight lines as a consequence of the 
linear-log  
(DEA case) and log-log (SDA case) representations we are adopting.  The 
global air  temperature anomalies are characterized by a pdf scaling 
coefficient $\delta_a=0.87\pm0.02$ and a standard deviation scaling 
coefficient $H_a=0.92\pm0.01$. The global sea  temperature anomalies 
are 
characterized by a pdf scaling coefficient $\delta_s=0.89\pm0.02$ and a 
standard deviation scaling coefficient $H_s=0.94\pm0.01$. The figures 
are 
plotted for a period $t=50$ months. For value of $t$ larger than 50, 
saturation effects due to the statistical property appear. The fitting 
is a 
kind of mean result obtained by averaging the results corresponding to 
fitting the first 20, 30 and 40 points. The upper time limit condition 
of $t = 50$ is dictated by the  fact 
that we are using N = 1680 data, a number with a  square root of 
about 40. The pictures show that for $t=50$ the standard deviation is 
almost 
40 times larger than the standard deviation at the first step of 
diffusion. 
It seems, therefore, that the statistics are rich enough to  get a 
satisfactory  pdf and consequently reliable scaling properties.

The high values of the exponents imply a strong persistence. This means 
that 
the temperature changes gradually month by month. The fact that the 
exponents 
stemming from  the sea temperature anomalies are higher than those 
produced 
by the air temperature anomalies means that the sea temperature 
anomalies are 
characterized by a persistence higher than that of the air temperature 
anomalies. This can be explained as an effect of the higher heat 
capacity of  
the water. These results confirm the results of Sec. 3, shown in Figs. 
4,  
where the fluctuations of the air temperature anomalies at short scale 
are 
stronger than those of the sea temperature anomalies. Finally, we note 
that 
the both exponents $H_a$, for  air,  and $H_s$, for sea, are larger 
than 
$\delta_a$ and $\delta_s$  respectively. This means that the pdf of the 
two 
diffusion processes is a little bit larger than a Gaussian 
distribution. The 
four exponents fulfill the L\'{e}vy Walk Diffusion relation 
(\ref{relHdelta34}) within the accuracy of our statistical analysis. 
This 
means that   
global air and sea temperature anomalies not only are characterized by 
L\'{e}vy statistics, but are a manifestation of the dynamic model 
illustrated 
in Section 4 A. This means that the alternated periods $\tau_i$ of high  
and 
low  temperature are distributed according an inverse power law with 
$\mu<3$.  According to Eqs. (\ref{scalrelH}) and (\ref{scalrelD}), we 
obtain $\mu_a=2.13 \pm 0.02$ and $\mu_s = 2.08 \pm 0.02$.  The results are summarized 
in 
Table III.

\section{Wavelet Multiresolution Diffusion Analysis}

In this section we introduce a method of analysis, based on the joint 
use of 
wavelet decomposition and the diffusion approach to scaling, the latter 
method resting on both DEA and SDA. This method turns out to be 
powerful, and 
we refer to it as Wavelet Multiresolution Diffusion Analysis, WMDA. 
Figs. 7 
and 8  show the SDA and the DEA of the global air (a) and sea (b)  
temperature applied to the residuals $R_j$, where, as in the earlier 
sections,  $j$ indicates the wavelet scale index.  Each residual contains all details at smaller scales. Therefore, the 
WMDA 
allows us to determine the diffusion spreading at each time scale, as 
stemming from the corresponding details, an important piece of 
information.   
The residuals $R_j$ are obtained by detrending the original data with 
the 
smooth curves $S_j$ obtained with the wavelet multiresolution analysis; 
see 
Eq. (\ref{decomwR}).    In each of the four figures there are nine 
curves. 
The figures (a) refer to air, the figures (b) refer to sea. Figs.  7 
refer to 
SDA and Figs.  8 to DEA. From top to bottom  the curves of these 
figures 
denote the original data (1), and  the residuals  (2) R9 (2) , (3) R8 
(3) 
,..., (9) R2(9) .  The curves of Figs. 7 and 8  are an illustration of 
WMDA 
and prove that this is a useful tool to study complexity of a dynamical 
system. Let us analyze them in detail.   

a- Figs. 7 and 8 look similar but are not identical. The curves 
afforded by  
DEA are  more detailed than those obtained by using SDA. This is 
because DEA 
is more sensitive to fluctuations than  standard deviation, which only 
measures the spreading of the diffusion trajectories.  

b- The straight lines correspond to the function $f_{SD}(t)=t^{0.5}$,  
in 
Figs. 7,  and to the function $f_{DE}(t)=0.5\ln(t)$,  in Figs. 8, and 
look 
straight due to the adoption of the  log-log (Figs. 7) and linear-log 
(Figs. 
8)  representations. These straight lines serve the purpose of 
signaling what 
would be the behavior 
of  an ordinary  Gaussian diffusion. According to the Mandelbrot 
interpretation \cite{2Mandelbrot}, the curve with slopes larger than 
that of 
these  straight lines indicate persistent diffusion or superdiffusion, 
and 
the curves with slopes smaller than that of  these straight lines 
indicate  
antipersistent diffusion or a subdiffusion region.  By comparing the  
curves 
of Figs. 7, resting on SDA, to the corresponding curves of Figs. 8, 
resting 
on DEA,  we see that the times at which the SDA curves cross the line 
$f_{SD}(t)=t^{0.5}$ are larger than the times at which the DEA curves 
cross 
the line $f_{DE}(t)=0.5\ln(t)$. This is a further proof that the 
dynamics 
behind the data cannot be Fractional Brownian Motion. This kind of 
diffusion 
process would imply   SDA and DEA crossing the lines denoting ordinary 
diffusion at the same time. The figures suggest that the diffusion is 
characterized by  a pdf whose tails  are more persistent than the 
Gaussian 
tails  of Fractional Brownian Motion. The analysis of Section 5C proved 
that 
the diffusion is anomalous and fulfills the L\'{e}vy Walk condition. In 
this 
section, with  the help of WMDA we  show that the data are 
characterized by 
an anomalous dynamics  at short as well as long time scales.

c- By comparing the  curves of  Figs. 7a and 8a to the corresponding 
curves 
of Figs. 7b and 8b,  we note the air temperature curves always have a 
slope 
smaller, even if slightly smaller, than the slope of  the corresponding 
sea 
temperature curves. This means that at each wavelet scale the higher 
heat 
capacity of the water makes the sea temperature data more persistent 
than the 
air ones.  This is related to the higher heat capacity of the water.

d- Figs. 7 and 8 show that each curve is characterized by one leading  
periodicity. The dynamical reasons for this property are easily 
accounted for 
by noticing that a given periodicity of the data causes a periodic 
convergence of distinct trajectories.  After an initial spreading, with 
a 
consequent increase of both variance and entropy, there are incomplete 
regressions of the initial condition.  Since each curve corresponds to 
a 
given scale, the observed processes of regression correspond to the 
leading 
periodicity of that temporal scale. Any residuals $R_j$ contains all 
details 
at smaller scales, and, as a consequence, the leading periodicity is 
not 
necessarily related to the wavelet temporal scale $\tau_j$ by simple 
relations. For example, the  R7 and R8 curves of Figs. 7a and 8a  show 
almost 
the same periodicity but $\tau_7$ is one half of $\tau_8$.  Moreover, 
the 
main periodicities do not coincide exactly for  the global air and sea 
temperature anomalies. The figures show that the year periodicity and 
its 
multiples have a strong effect until the residual R6. A periodicity of 
17-20 
years dominates the R7 and R8 residuals. Finally, a periodicity of 
57-59 
years characterizes the R9 rest. 
Figs. 7 and 8 show that WMDA may be an interesting complement to the 
spectral 
density analysis of Figs. 2 because it shows the main periodicity and 
its 
contribution to the information that characterizes that scale level.  

e- Figs. 7 and 8 show that both entropy and standard deviation of any 
residual converges to a horizontal line. This is due to the detrending 
of the 
smooth part of the data, $S_j$, which makes the hidden periodicities 
show up. 
At the same time, these hidden periodicities imply the deterministic 
nature 
of the signal and consequently yield entropy saturation. In other 
words,  the 
fluctuations $\xi_i$ of the residuals data $R_j$ can generate  
trajectories 
with only a limited spreading.    The height of the horizontal lines 
measures 
the maximum spreading (in the case of SDA, Figs. 7) and the information 
or 
entropy (in the case of DEA, Figs. 8) for each time scale. It is 
interesting 
to notice that the shorter the time scale the faster the transition to 
saturation. This means that the role of periodicities becomes more and 
more 
important as we decrease the time scale. 

Table IV summarizes some of the information contained in  Figs. 7 and 
8, and 
can help the reader to understand the balance of the results of this 
paper 
that will be given in Sec. 6.

Finally, the reader may wonder what happens if we detrend one of the 
wavelet 
details from the original data. 
 In our analysis we observe that the scaling properties coincide, 
within the 
limit of the analysis statistical accuracy,  with those of Section 4 C  
for  
$t<50$. This applies to detrending from  the original data any detail 
between 
D1 and D10. This means that the wavelet smooth curves S determine the 
scaling 
because they determine the persistent properties of the signal. The 
scaling 
for  $t<50$ is not conditioned by any cycle in the data.  If  we 
consider 
interval larger than $t=50$ we observe the largest discrepancy by 
detrending 
the detail D7. Figs 9a (air) and 9b (sea) show the results. The 
detrended 
data have a lower entropy for $30<t<200$. This entropy seems to scale 
with 
$\delta$ close to 0.5 until $t=600$, like a Brownian diffusion. 
However, the 
lack of sufficiently rich statistics in that region does not allow us 
to get 
any conclusion about the real value of the scaling. The SDA, instead, 
does 
not detect this difference, Figs 9c and 9d; the entropic analysis turns 
out 
to be more sensitive than the variance analysis.    The increasing of 
entropy 
of the original data for $30<t<200$ have a non-stationary origin due to 
a 
cycle that is detrended by the detail D7. In fact, detail D7 
corresponds to 
the wavelet time scale of 128 months, that is, 10-11 years, and it 
contains 
the important 10-11 years solar cycle.

\section{Conclusion}

The key results of these paper can be summarized as follows:

(i) \emph{Understanding the various aspects of a global climate 
system}.
 Global climate is a complex system in which many factors affect one 
another 
and without a detailed look into the behavior of the Earth different 
temperature systems and their correlations, any attempt to understand 
it will 
be difficult at best.  For example,  understanding these behaviors is 
important to comprehend local and global climate changes, as well as, 
to 
attempt to rebuild past climate more accurately \cite{MBH1999}.  
In fact, our analysis shows that the differences between the behavior 
of the 
data on the shorter time scale lends itself to caution when performing 
these 
reconstructions.  The understanding and reconstruction of climate to 
times 
previous to instrumental records will allow us to gain some knowledge 
of the 
long term behavior of future climate \cite{high}.    We show, in fact, 
that 
no significant correlation exists for detail up to time intervals of 16 
months. For larger time intervals this correlation becomes visible.  
Our 
analysis has shown the differences in the fundamental types of 
statistical 
behavior of both regions of the Earth.  This knowledge could be used to 
further our ability to reconstruct past climate in an attempt to better 
understand our dynamic global environment.

(ii) \emph{Scaling and its relevance for the air and sea temperature 
anomalies}.
The joint use of DEA and SDA allowed us to establish better the nature 
of the 
signals, as well as their  scaling. The data seems to have a slight 
L\'{e}vy 
component. The higher scaling of sea signal is interpreted as a 
consequence 
of the higher thermal capacity of sea water.

(iii) \emph{Periodicity effects on scaling}.
We proved that scaling detected by using the DEA is not affected by 
periodicities. The smooth curves are responsible for scaling and  
scaling is 
not influenced 
by cycles. The smooth curves are responsible for a steady increase of 
entropy 
of the diffusion. 

(iv) \emph{Residuals and periodicities}.
We showed that periodicities emerge at the level of  residuals, namely 
the 
portions of the signals obtained detrending the smooth parts. The 
diffusion 
entropy of residuals saturates, thereby implying that after a given 
time 
there is no further information increase.
We can therefore, conclude that there should be no concern about a 
possible 
influence of periodicities on scaling. The scaling found is a genuine 
property and we can freely adopt it as an indicator of correlations. 
The sea 
temperature data, yielding a higher scaling, imply a larger correlation 
and 
the statistical analysis of this paper makes compelling this important 
conclusion. 

(v) \emph{Details and non-stationary effect}.
We show that detail D7, that contains the important 10-11 year solar 
cycle,  
causes a non-stationary effect that is detected by DEA but not by SDA. 
This 
shows that DEA has a higher sensitivity. 

(vi) \emph{Joint use of entropy and decomposition}. 
We think the benefit of the joint use of DEA and wavelets is evident. 
The 
wavelet decomposition generates a set of new time series, corresponding 
to 
tuning the wavelet microscope to a given time scale, and the DEA 
establishes 
the information of these components, and makes it evident why 
periodicities 
set an upper limit on entropy increase. 

It is interesting to point out that  the earlier analysis of 
temperature 
anomalies\cite{nicolis,gimeno} has been done using deterministic chaos 
and 
the evaluation of the Lyapunov coefficient, and so the Kolmogorov-Sinai 
(KS) 
entropy. Our research work seems to support the dynamical model of 
Ref.\cite{dea7} , which is a form of L\'{e}vy walk. This means a 
connection 
with turbulence and intermittence. The values of $\mu$ emerging from 
this 
analysis are very close to $\mu = 2$ and thus to the border with the 
non-stationary dominion \cite{massi}. It is known that the KS entropy 
vanishes for $\mu \leq 2$. The authors of Ref.\cite{gimeno} seem to 
rest on a 
condition of vanishing KS entropy to address the intriguing problem of 
climate predictability. We think therefore that the joint use of DEA 
and 
Compression algorithms\cite{dea6}, applied after
the wise detrending method illustrated in this paper, might contribute  
further  processes towards the ambitious
goal of predictability. 

--------\\
{ {\large \bf Acknowledgment:}}\\
P.G. gratefully acknowledges the financial support received from the 
Army Research Office through Grant DAAD 19-02-0037 and N.S. thanks the 
Army 
Research Office for support under grant DAAG5598D0002.

%%%%%%%%%%%%%%%%%%%%%%%%%%%%%%%

%%%%%%%%%%%%%%%%%%%%%%%%%%%%%%%%
\newpage

\onecolumn

\begin{figure}
Figure 1\\
\epsfig{file=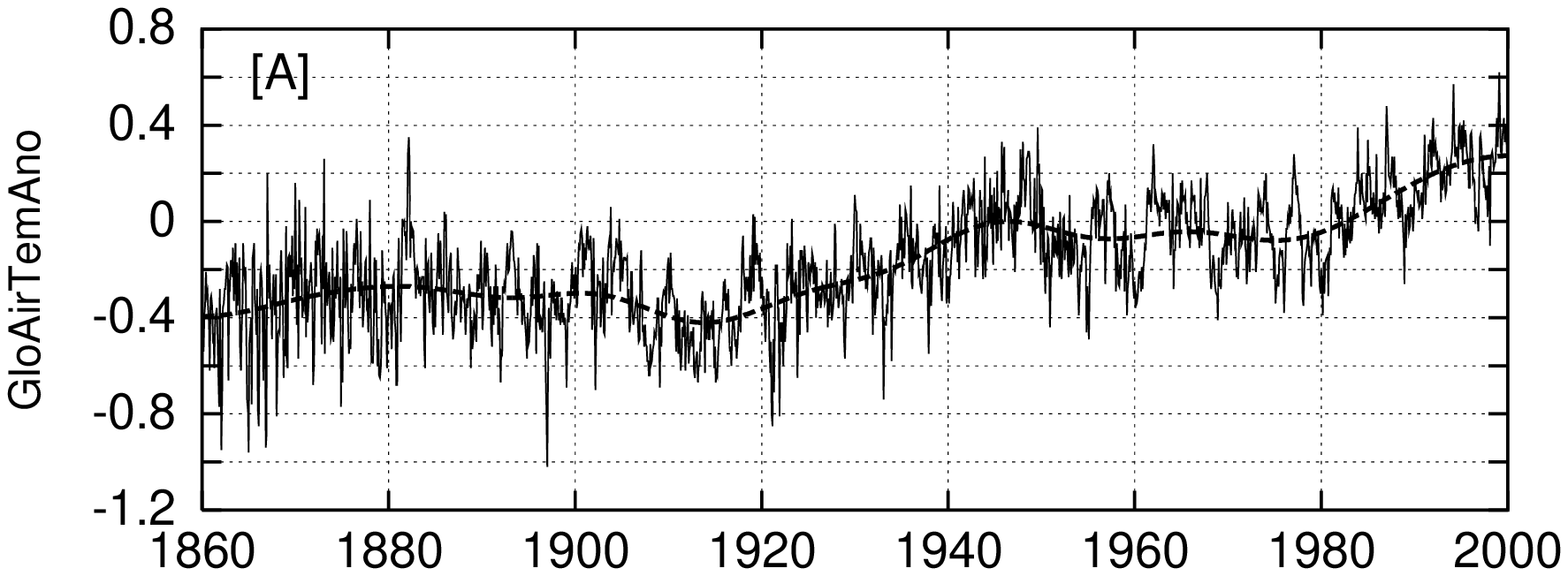,height=17cm,width=8cm,angle=-90}\\
\epsfig{file=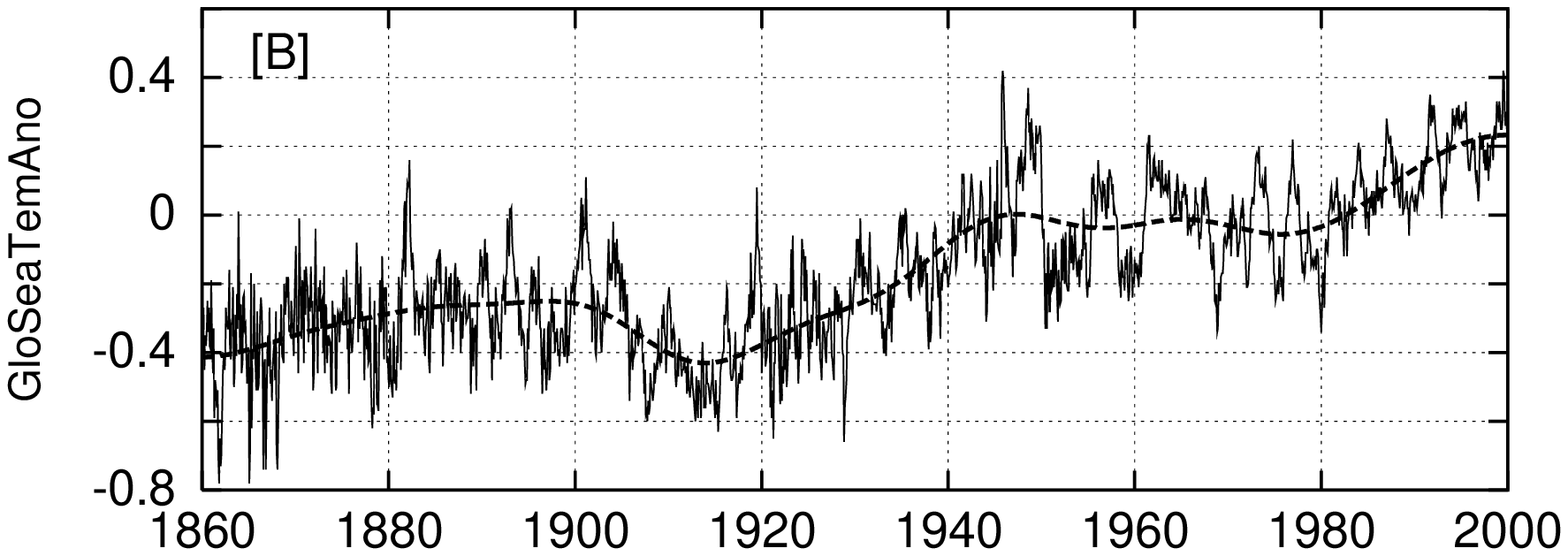,height=17cm,width=8cm,angle=-90}

\caption{Global air and sea temperature anomalies in Celsius degree (years: 1860-2000). 
The 
dashed lines are the wavelet multiresolution smooth curves S7.}

\end{figure}

\newpage

\begin{figure}
Figure 2\\
\epsfig{file=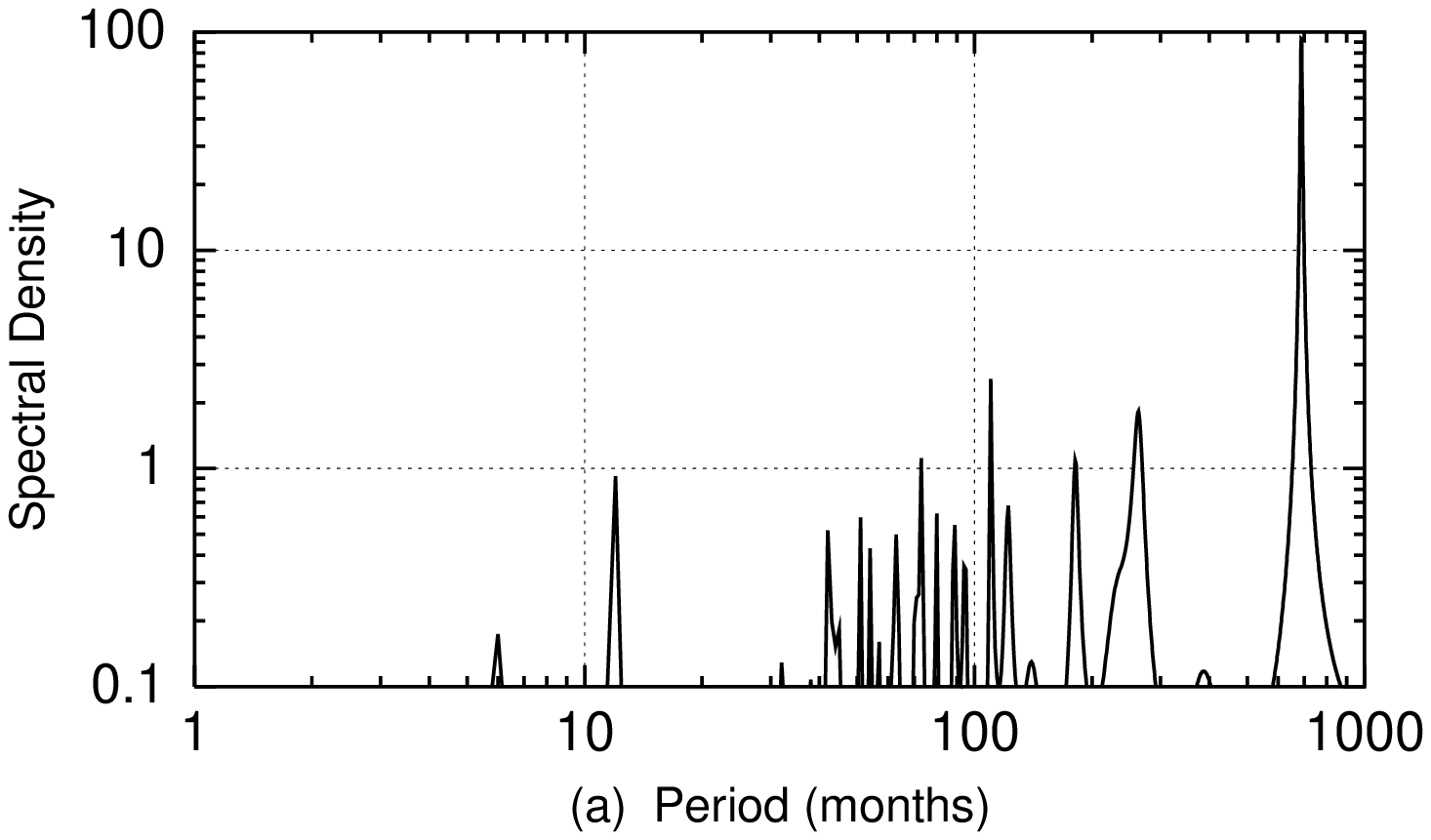,height=17cm,width=8cm,angle=-90}\\

\epsfig{file=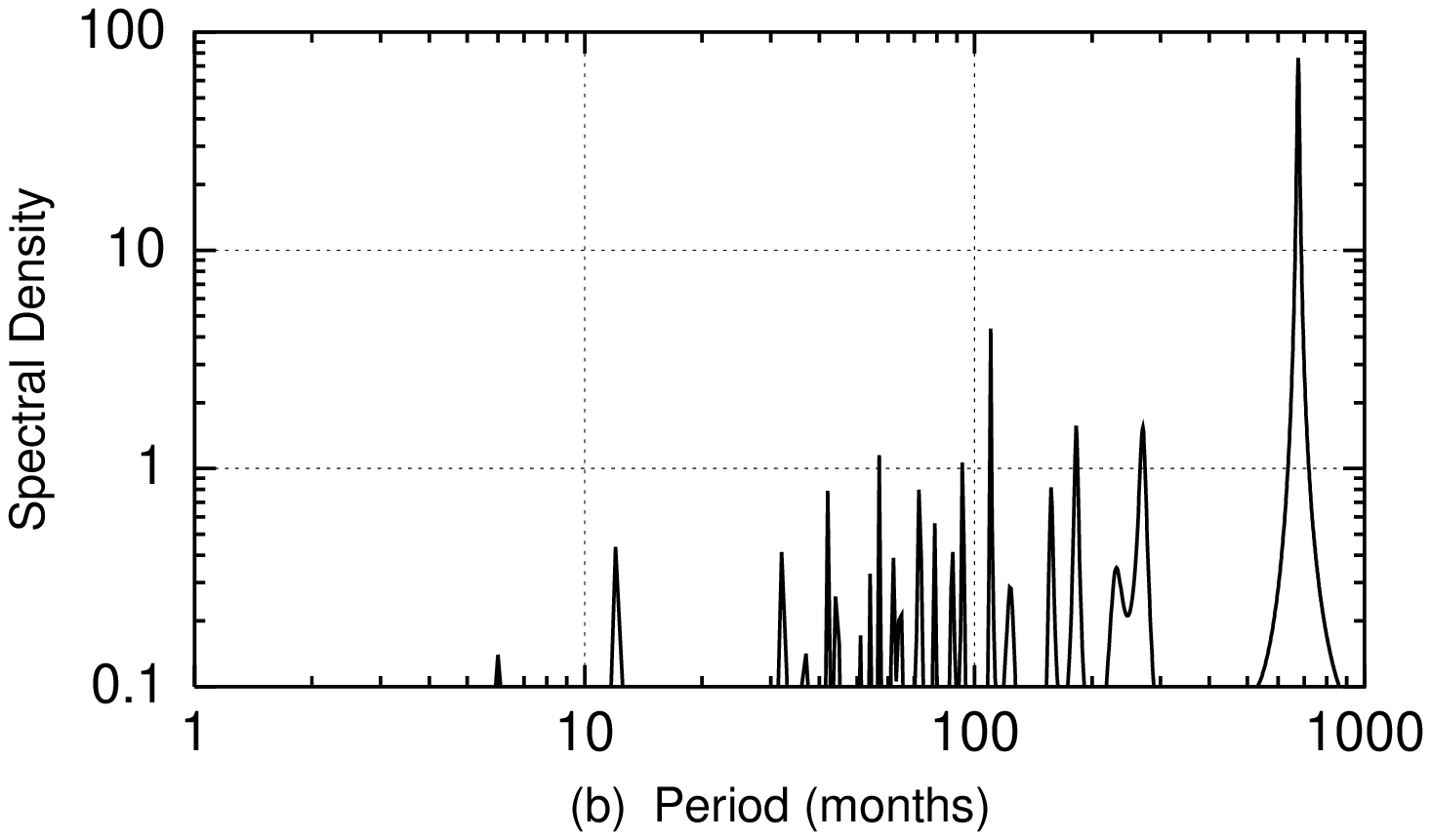,height=17cm,width=8cm,angle=-90}

\caption{Global air (a) and sea (b) temperature anomalies spectral 
density 
analysis against the period. In Table I there are some of the main 
periods.}

\end{figure}

\begin{table}
  \begin{tabular}{|c|c|c|c|c|c|c|c|c|c|c|c|}

 Air         &    1     & &  3.5  & 4.25 & 6.1&  & 9.2 & 10.2 & 15.1 & 
21.9 & 
55.7   \\ \hline
Sea         &    1   & 2.7 & 3.5  &  4.75 & 6 & 7.75 & 9.2 & 12.7  & 
15.2 & 
22.5 & 56.3   

\end{tabular} 
\caption{Main periods present in the global air and sea temperature 
anomalies.  The values are in years.}
\end{table}

\newpage

\begin{figure}
Figure 3\\
\epsfig{file=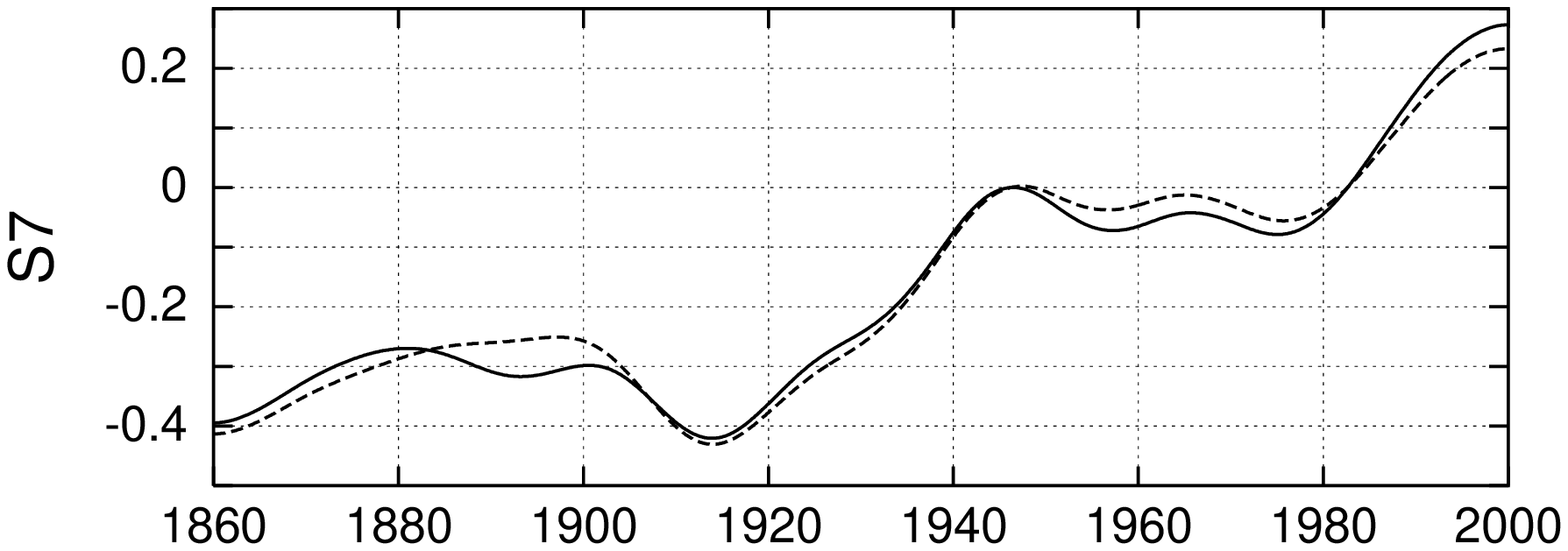,height=8.9cm,width=3.5cm,angle=-90}
\epsfig{file=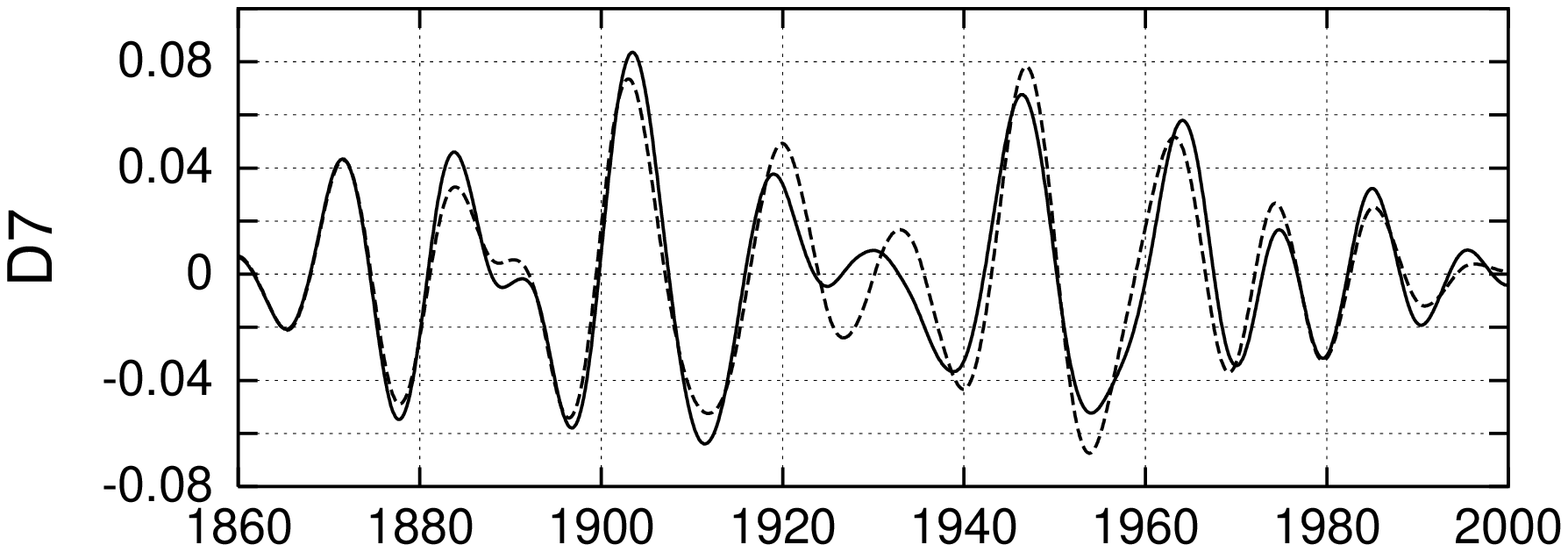,height=8.9cm,width=3.5cm,angle=-90} \\
\epsfig{file=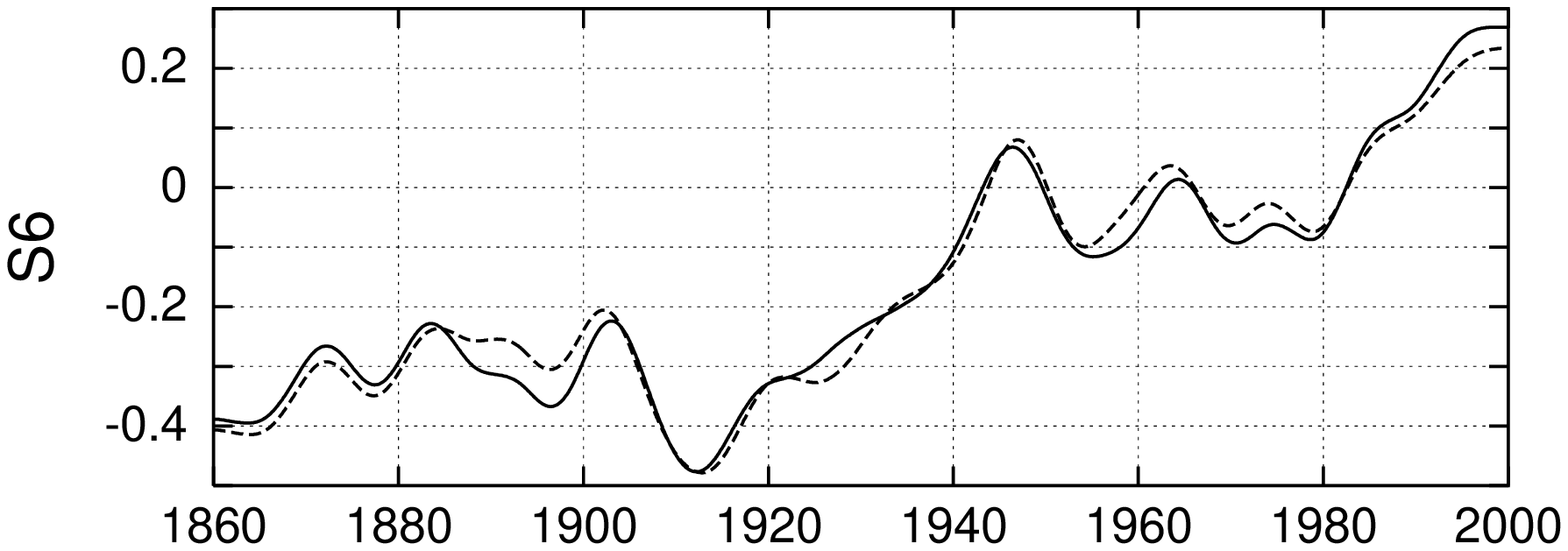,height=8.9cm,width=3.5cm,angle=-90}
\epsfig{file=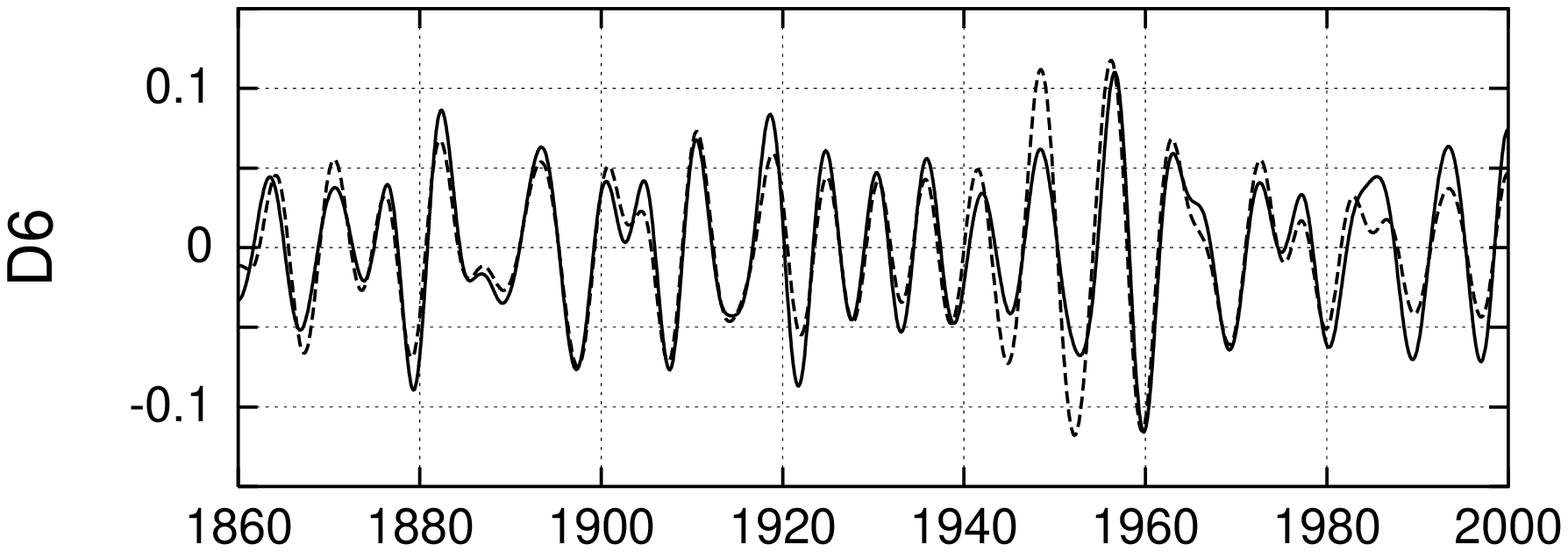,height=8.9cm,width=3.5cm,angle=-90} \\
\epsfig{file=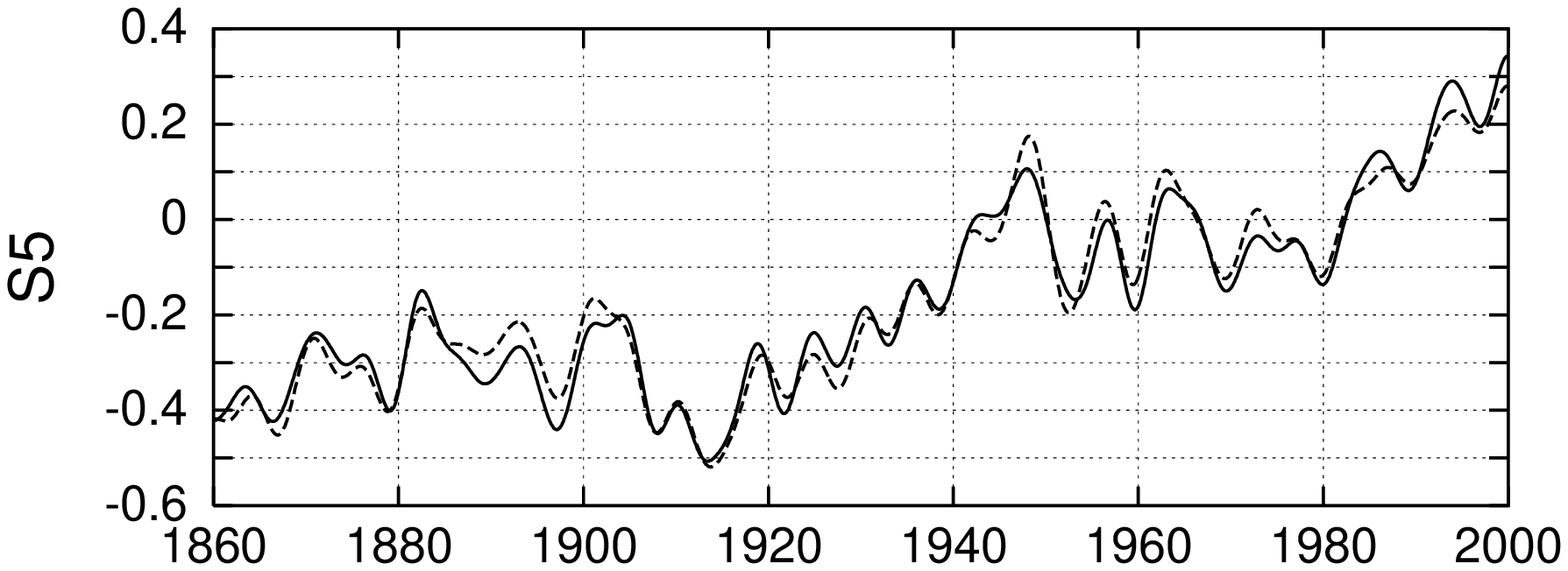,height=8.9cm,width=3.5cm,angle=-90}
\epsfig{file=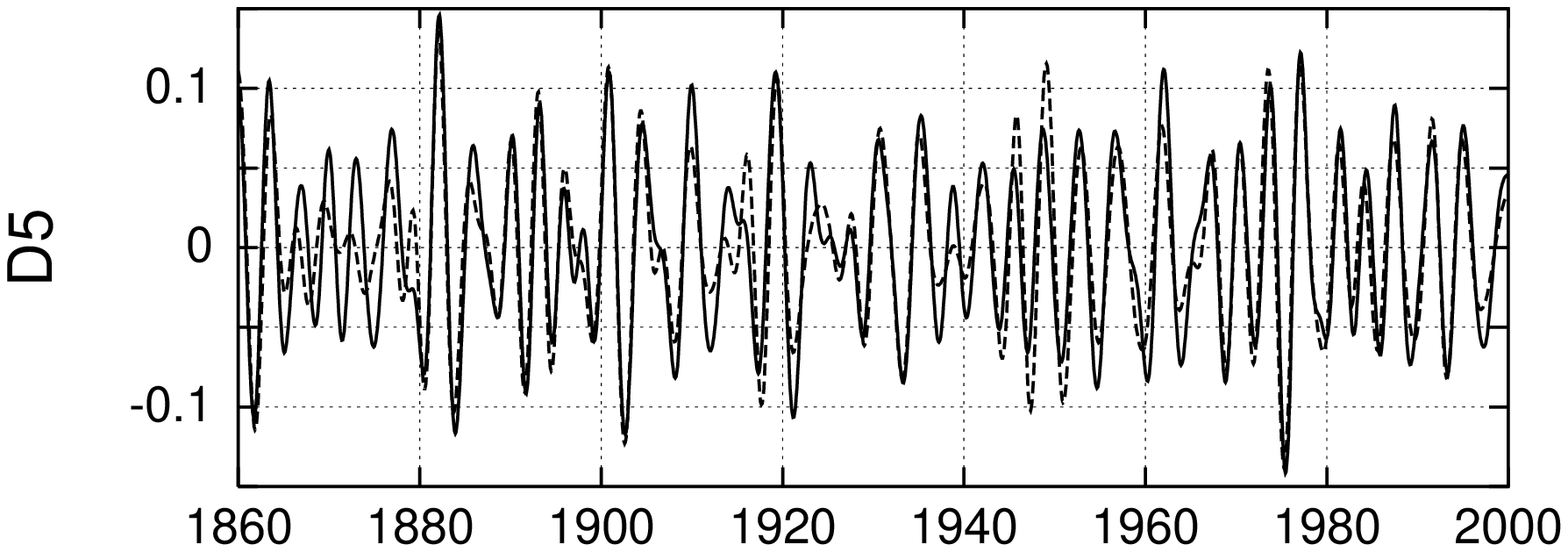,height=8.9cm,width=3.5cm,angle=-90} \\
\epsfig{file=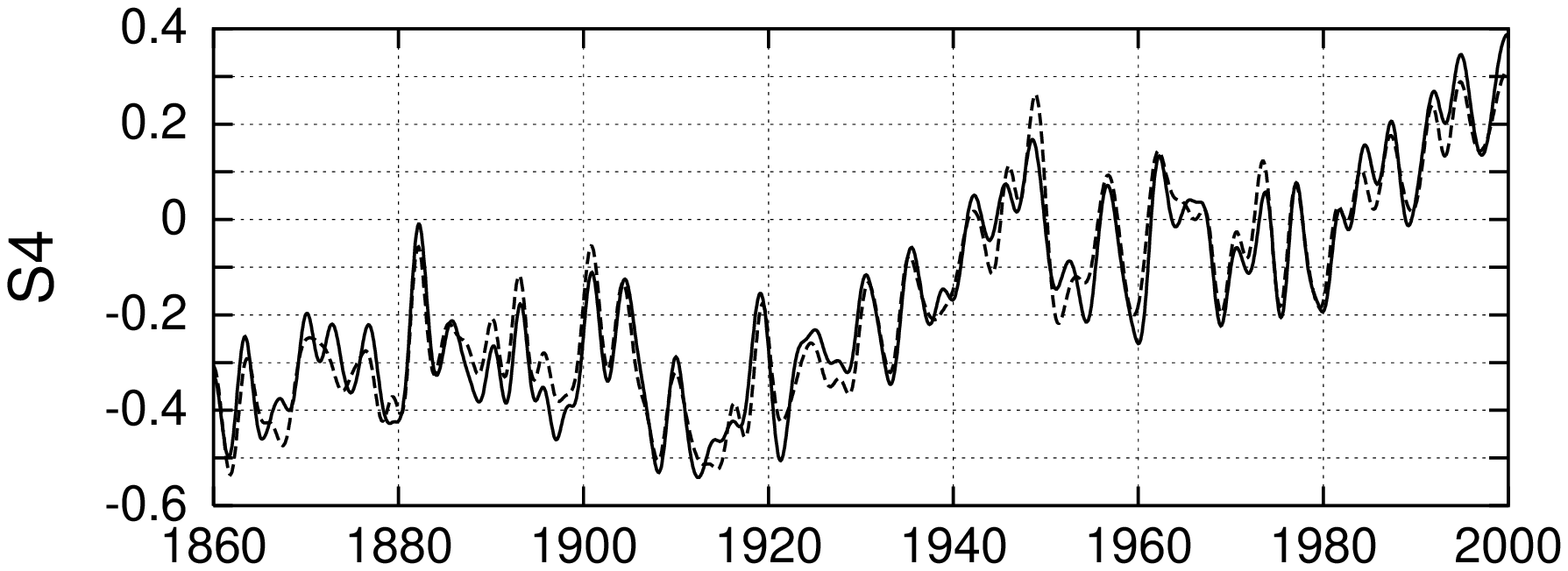,height=8.9cm,width=3.5cm,angle=-90}
\epsfig{file=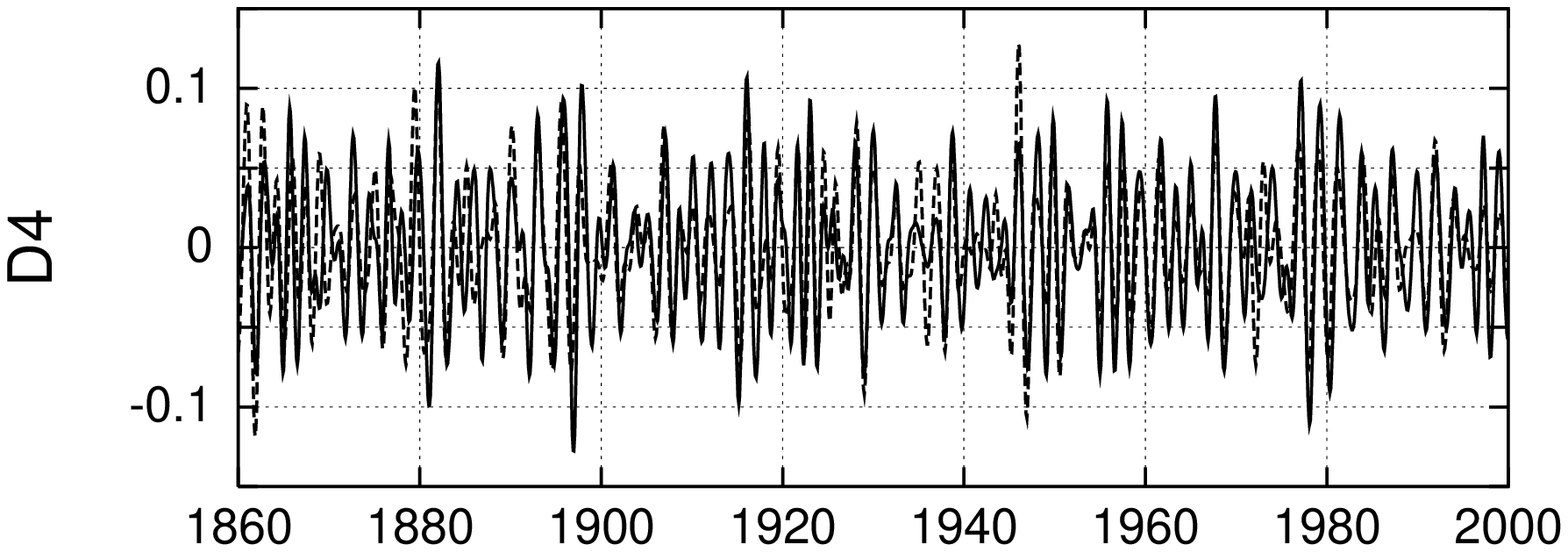,height=8.9cm,width=3.5cm,angle=-90}

\caption{Wavelet Multiresolution Analysis of the global air (solid 
lines) and 
sea (dashed lines) temperature anomalies in Celsius degree. The figures show 
the 
smooth curves S7, S6, S5, S4 and the details curves D7, D6, D5 and D4 
that 
are associated to the scales of 256, 128, 64 and  32 months. }

\end{figure}

\newpage

\begin{figure}

Figure 4\\
\epsfig{file=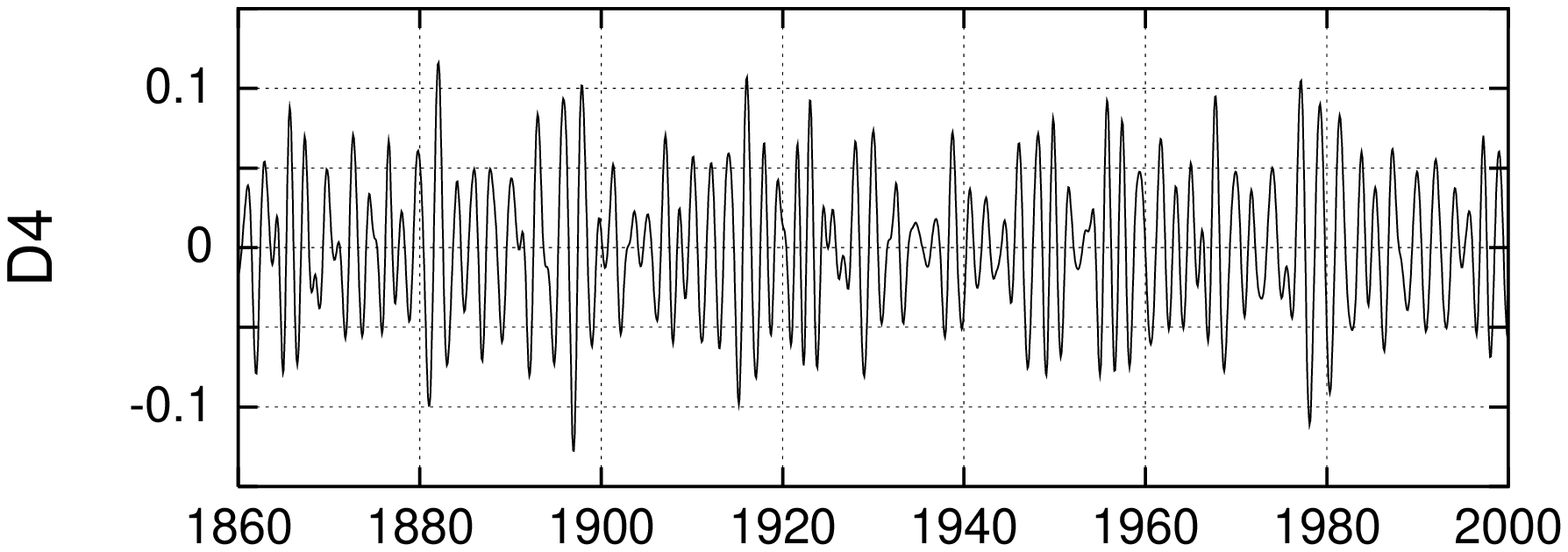,height=8.9cm,width=3.5cm,angle=-90}
\epsfig{file=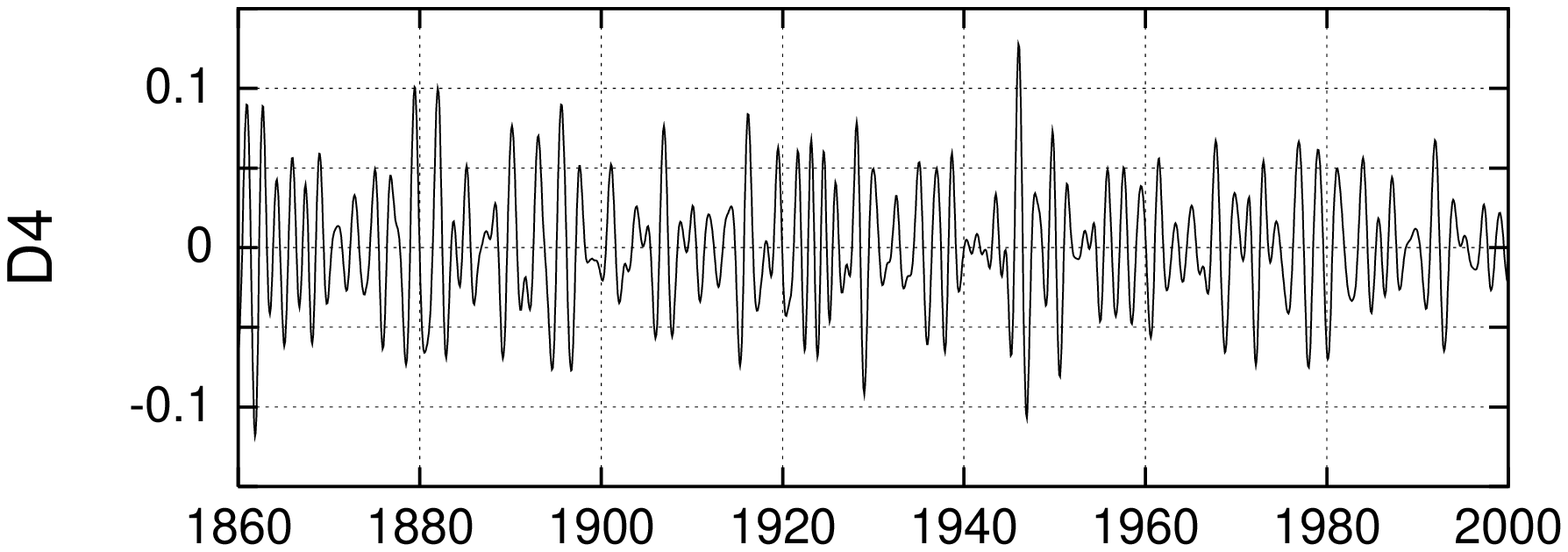,height=8.9cm,width=3.5cm,angle=-90} \\
\epsfig{file=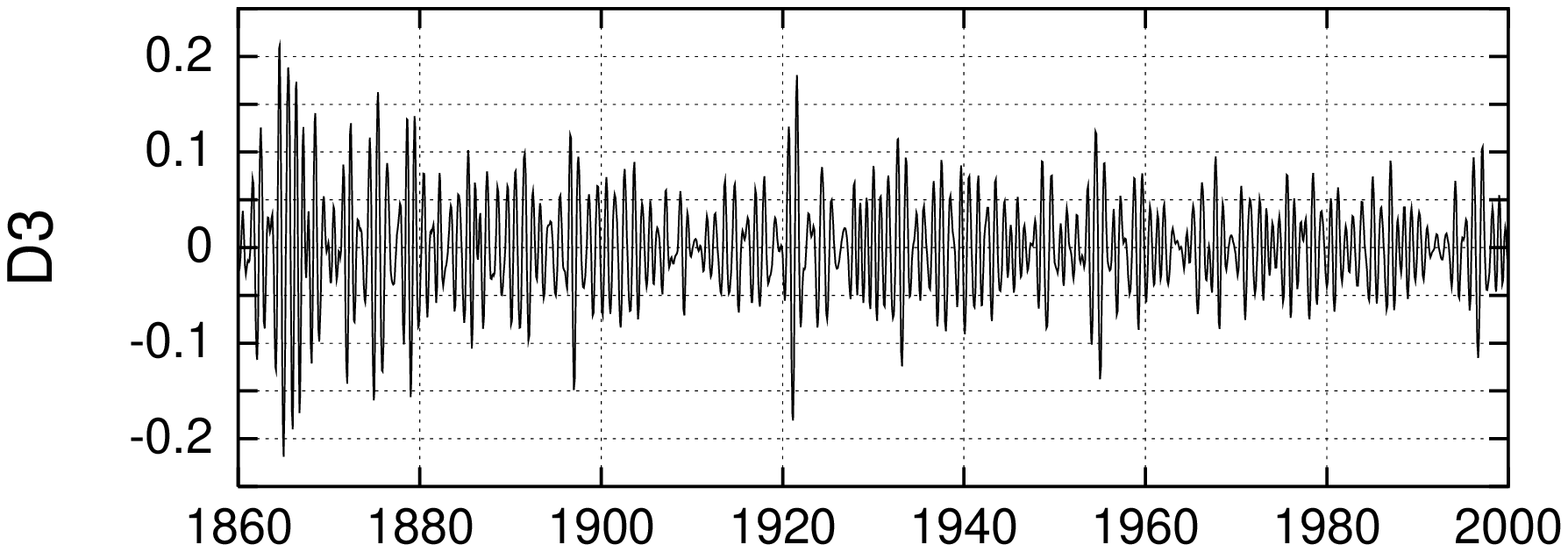,height=8.9cm,width=3.5cm,angle=-90}
\epsfig{file=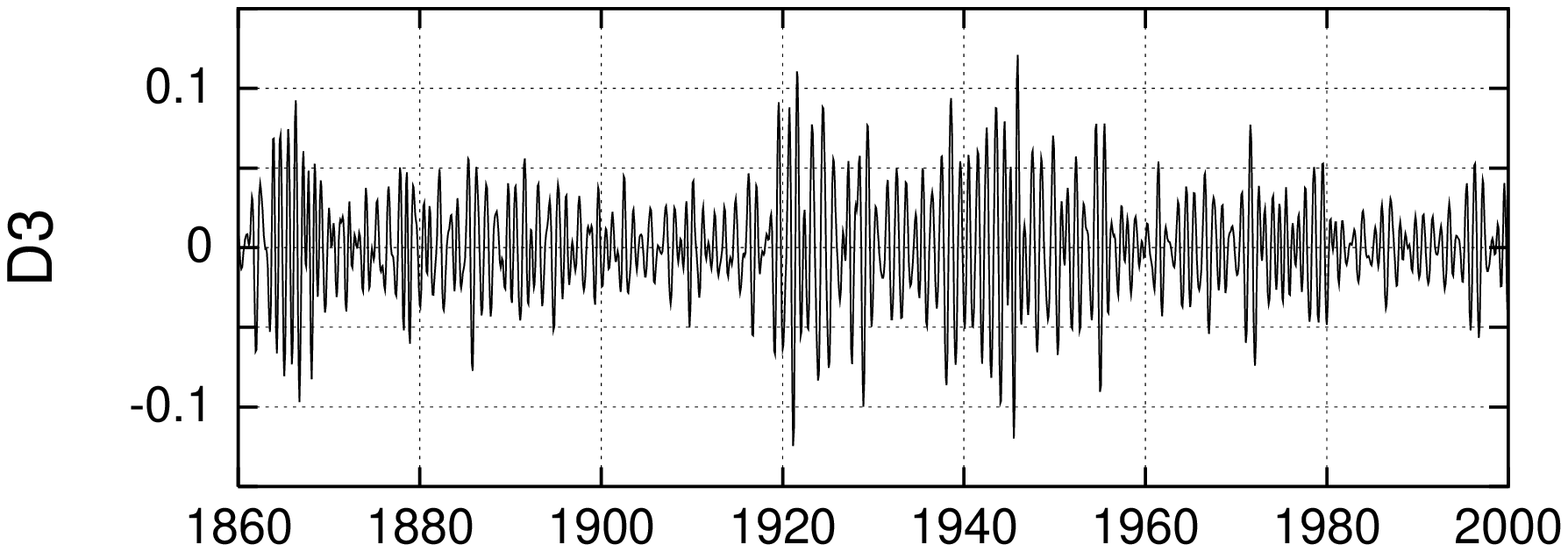,height=8.9cm,width=3.5cm,angle=-90} \\
\epsfig{file=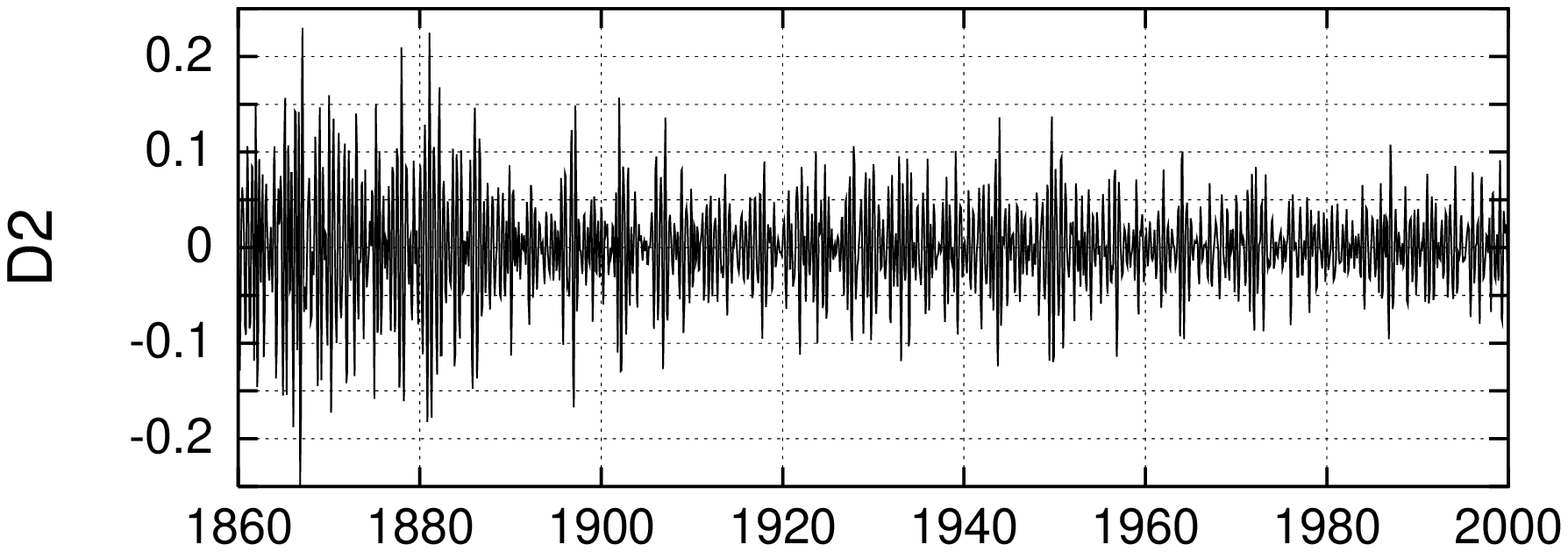,height=8.9cm,width=3.5cm,angle=-90}
\epsfig{file=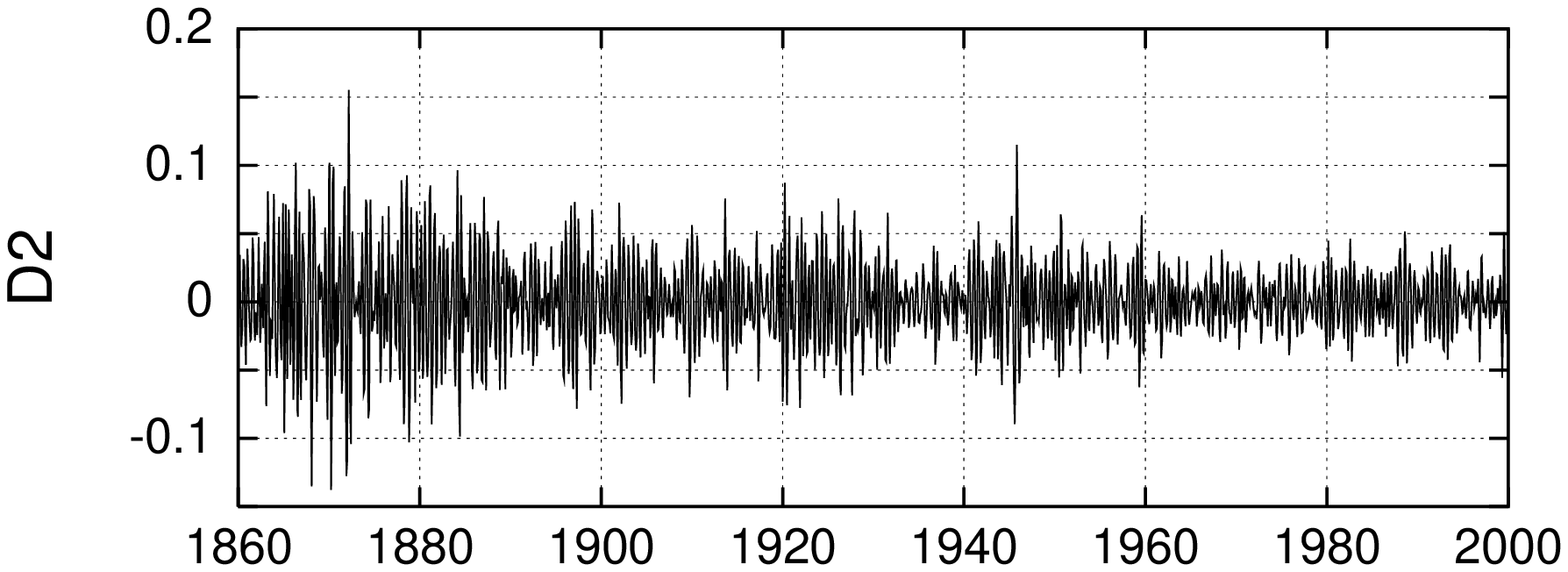,height=8.9cm,width=3.5cm,angle=-90} \\
\epsfig{file=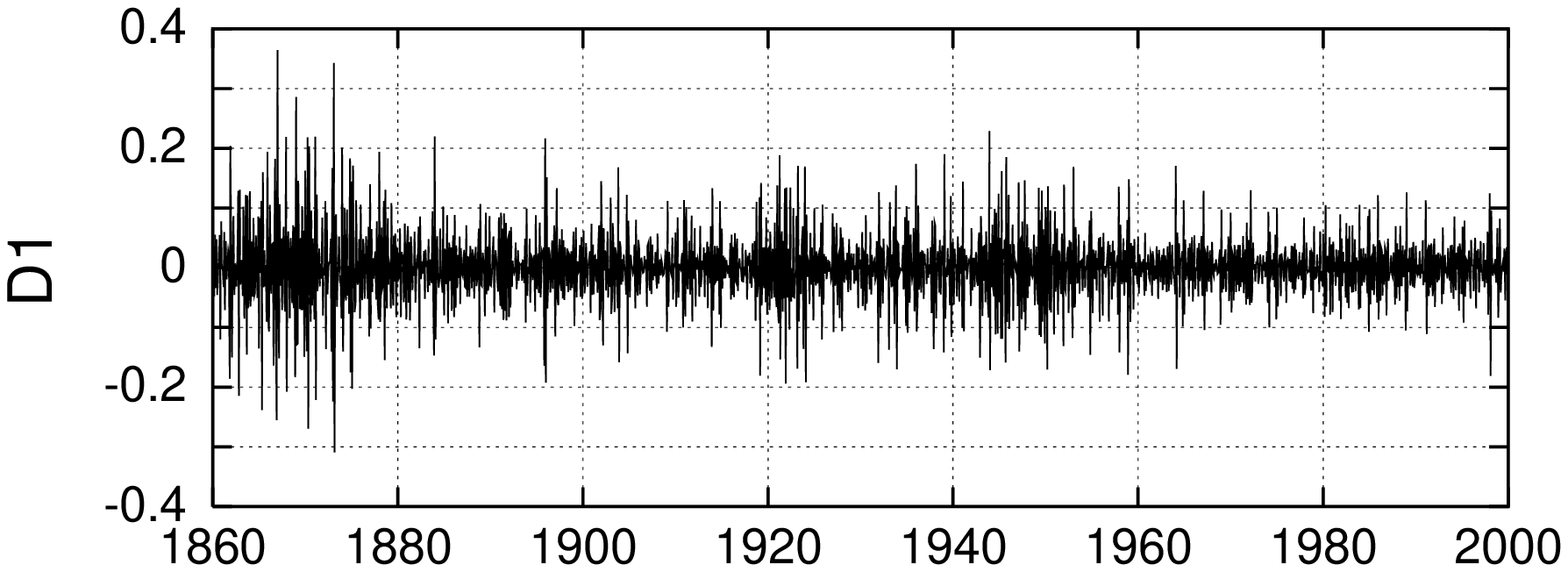,height=8.9cm,width=3.5cm,angle=-90}
\epsfig{file=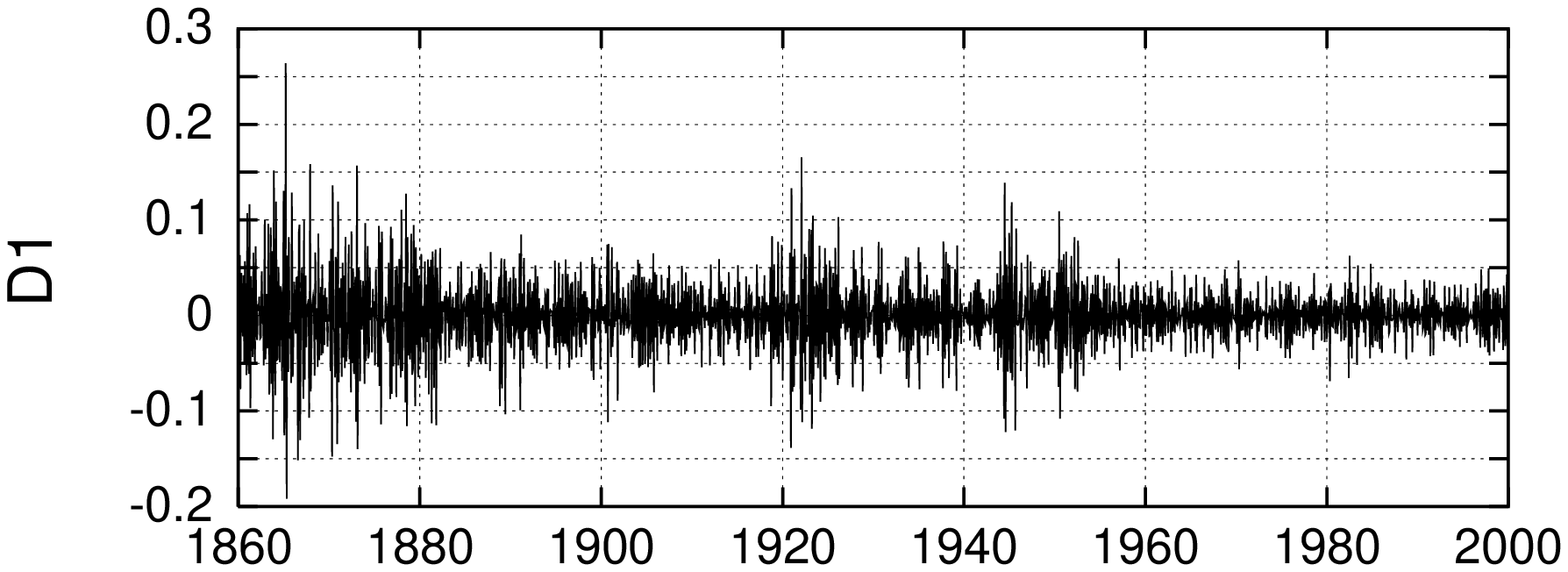,height=8.9cm,width=3.5cm,angle=-90}

\caption{Wavelet Multiresolution Analysis of the global air (left) and 
sea 
(right)  temperature anomalies  in Celsius degree. The figures show the details D4, D3, 
D2, D1 
that are associated to the scales of 16, 8, 4, 2 months.}

\end{figure}

\newpage

\begin{figure}
Figure 5\\
\epsfig{file=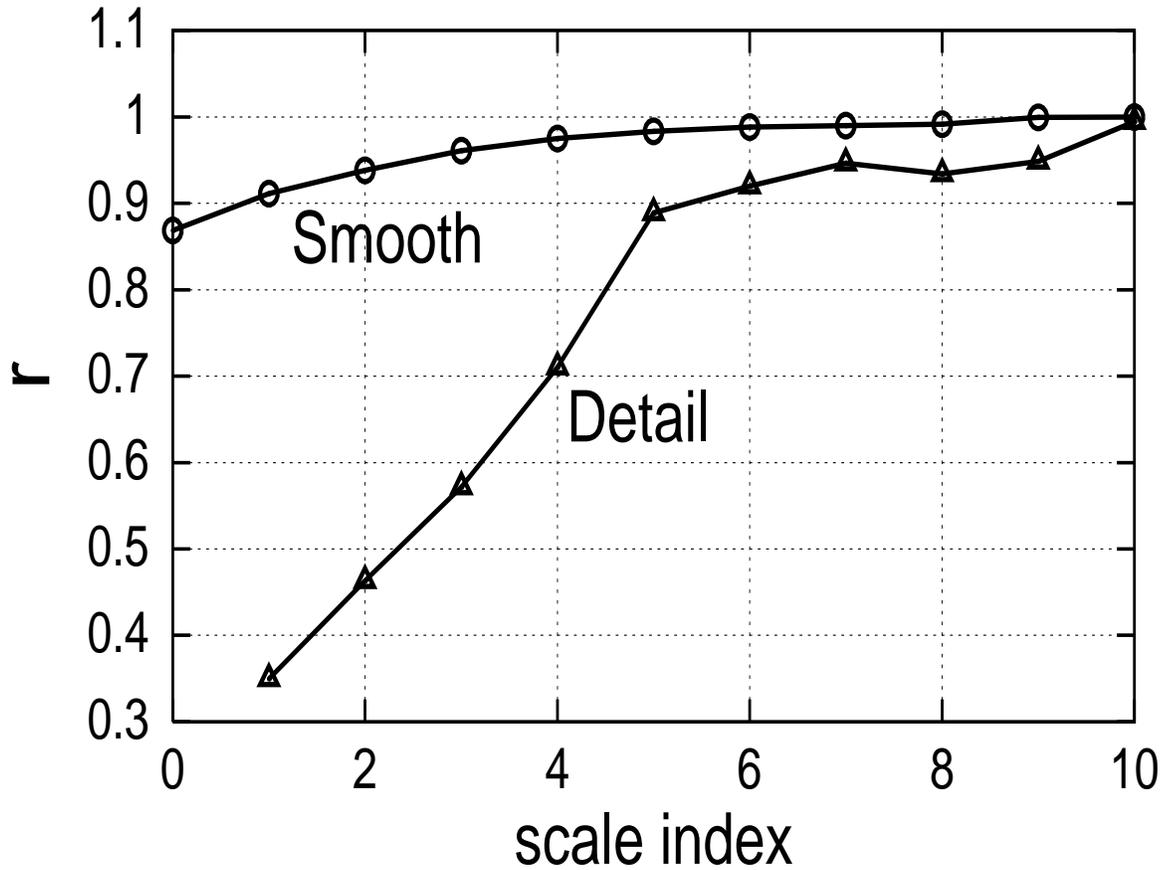,height=16cm,width=12cm,angle=-90}

\caption{Correlation coefficient $r$ between the global air and sea 
temperature anomalies against the wavelet scale index. The top curve 
denotes 
the correlation coefficient between  the wavelet smooth curves, from S0 
to 
S10. The smooth curves S0 refer to the original data without any 
filtering. 
The bottom curve denotes the correlation coefficient between the 
wavelet 
detail curves from D1 to D10. See Table II for the value of $r$.}

\end{figure}

\begin{table}
  \begin{tabular}{|c|c|c|c|c|c|c|c|c|c|c|c|}

  index             & 0 & 1& 2  & 3& 4& 5& 6& 7 & 8& 9 & 10  \\ \hline  
Smooth         & 0.87        & 0.91&  0.94  & 0.96 & 0.97& 0.98 & 0.99 
& 0.99 
& 0.99 & 1.00 & 1.00   \\ \hline
Detail         &       & 0.35 & 0.46  &  0.57 & 0.71 & 0.89 & 0.92 & 
0.95  & 
0.93 & 0.95 & 0.99  

  \end{tabular} 

\caption{Correlation coefficient $r$ between the global air and sea 
temperature anomalies against the wavelet scale index. The value 
$r=0.87$ of 
S0 is the correlation coefficient between the global air and sea 
temperature 
anomalies without any filtering.}

\end{table}

\newpage

\begin{figure}

Figure 6\\
\epsfig{file=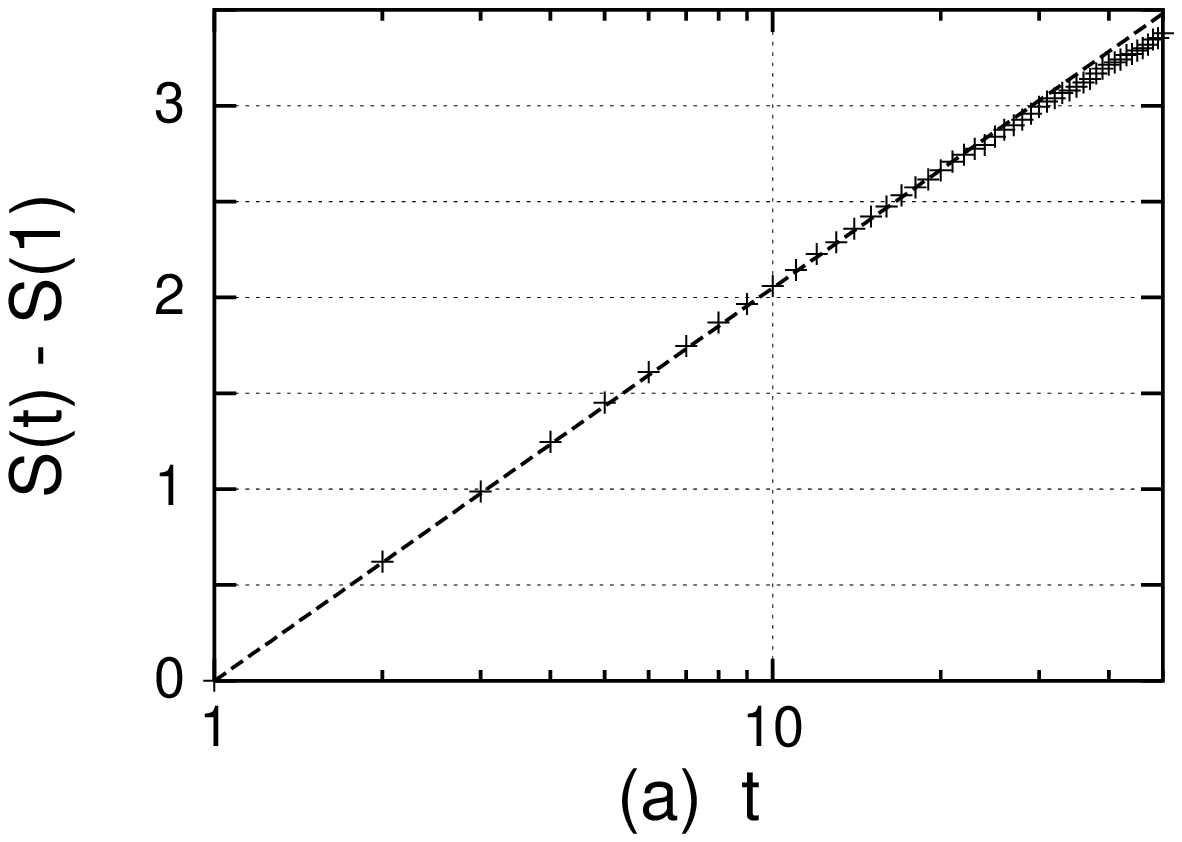,height=8.9cm,width=6cm,angle=-90}
\epsfig{file=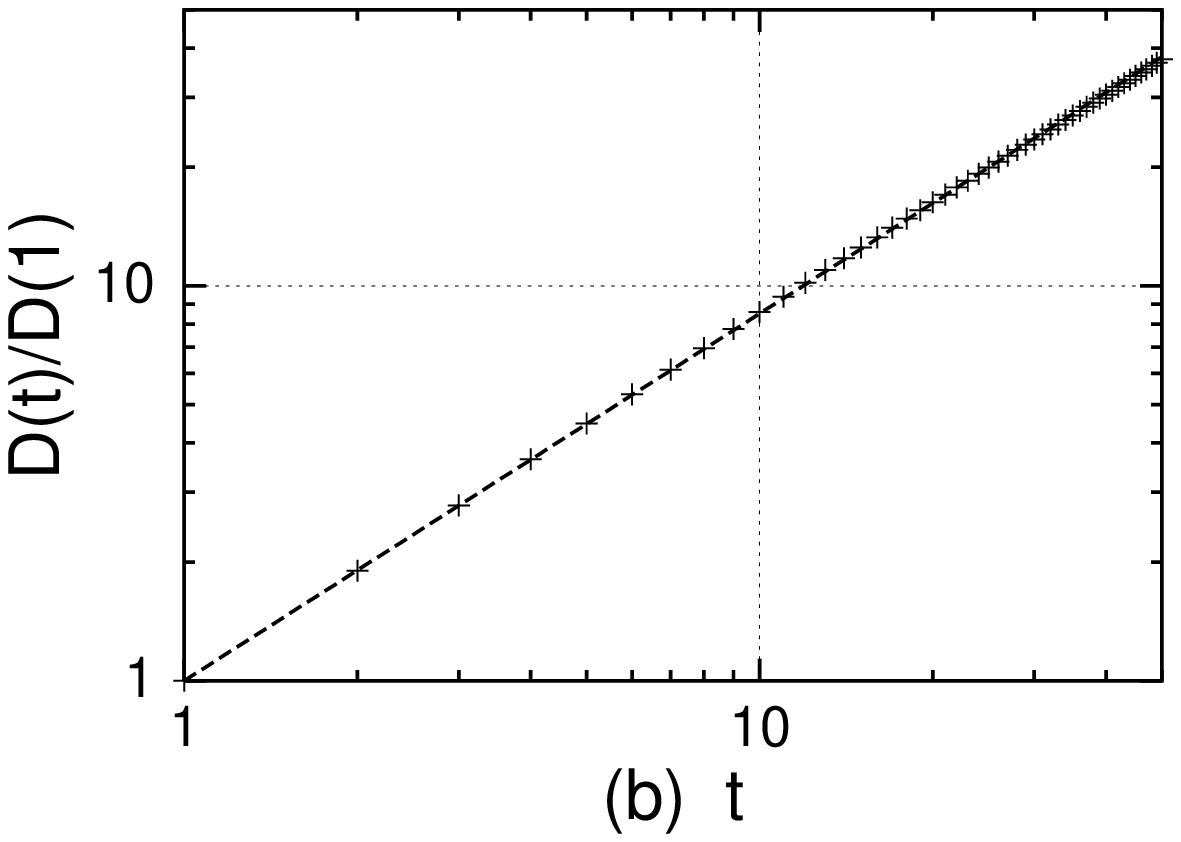,height=8.9cm,width=6cm,angle=-90} \\
\epsfig{file=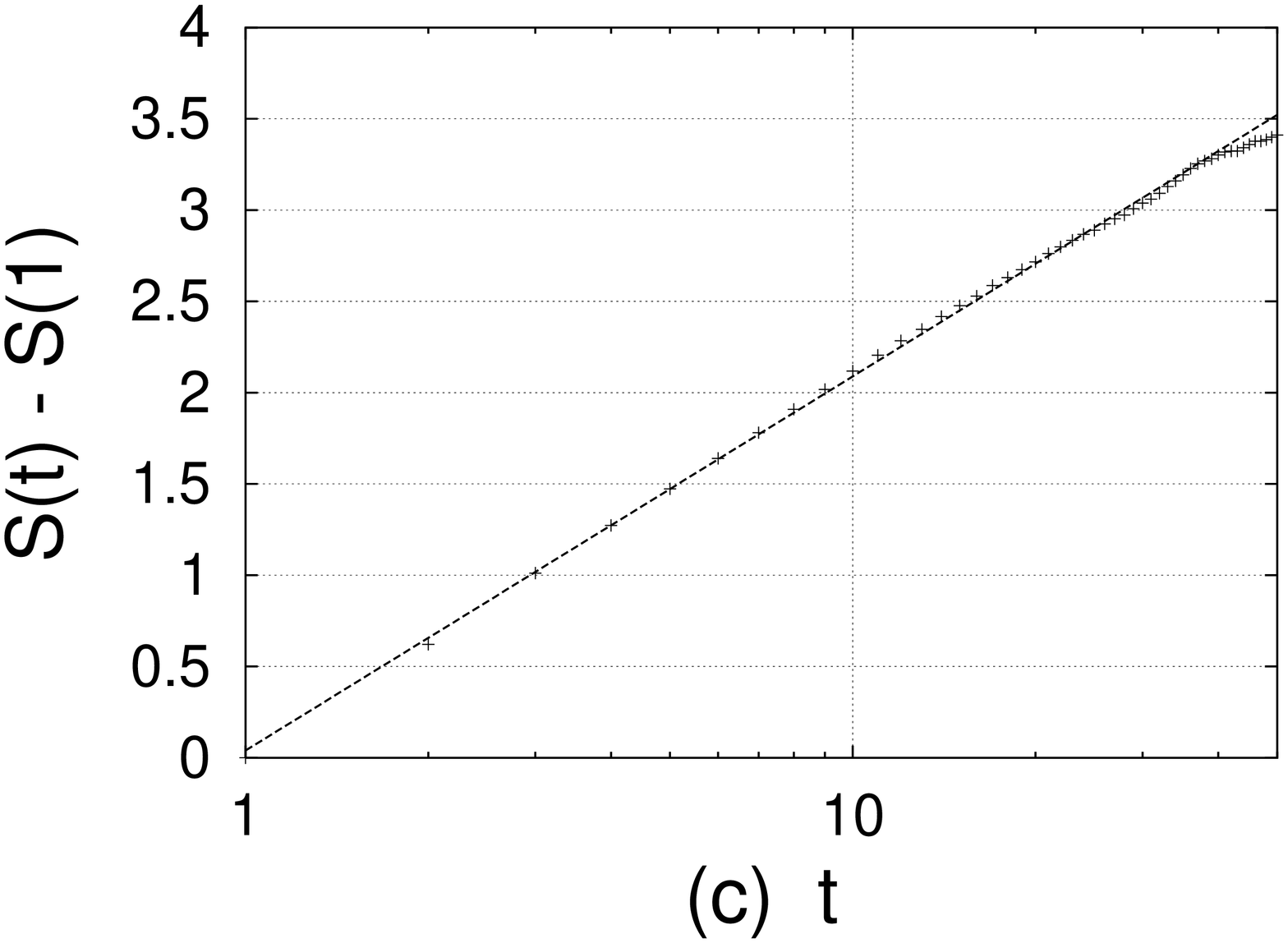,height=8.9cm,width=6cm,angle=-90}
\epsfig{file=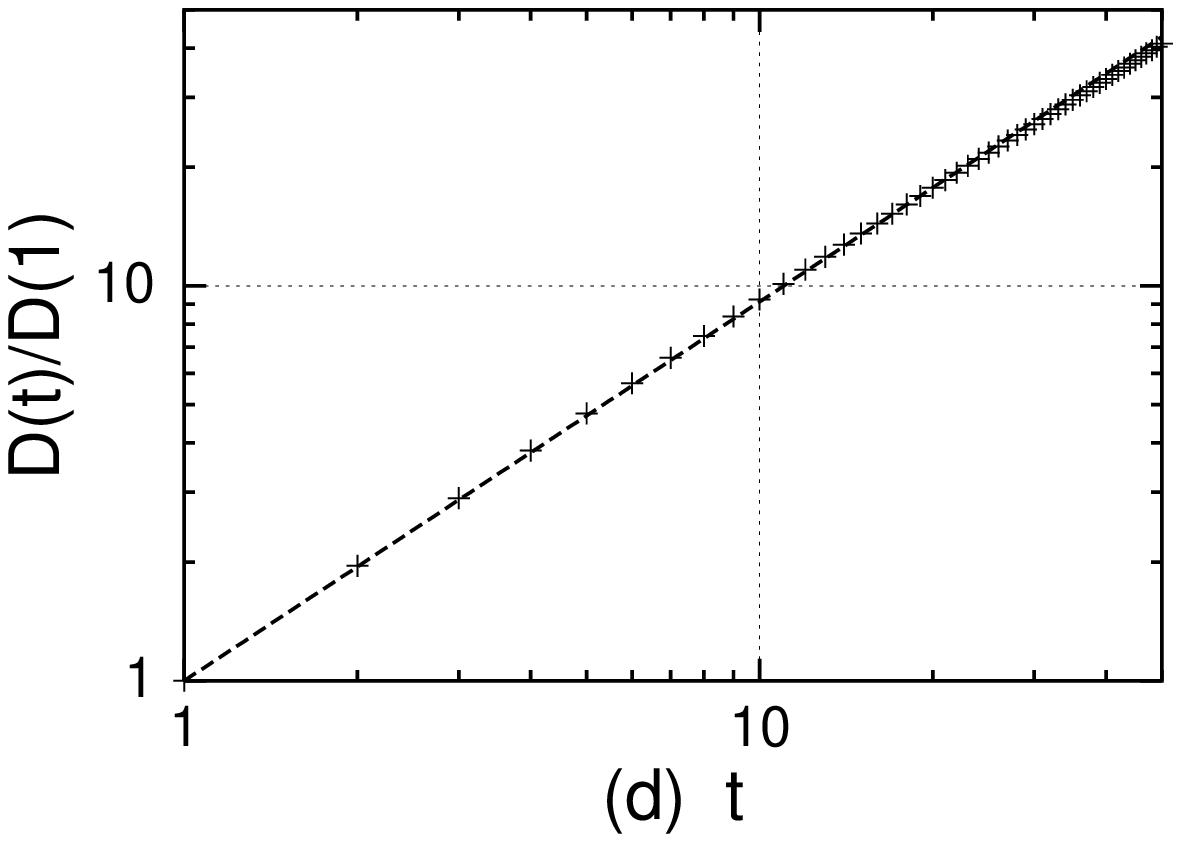,height=8.9cm,width=6cm,angle=-90}

\caption{DEA (Figs. a and c) and SDA (Figs. b and d) of the global air 
(Figs. 
a and b) and sea (Figs. c and d)  temperature anomalies. The straight 
lines 
correspond to functions of the type $f_{DE}(t)=\delta ~\ln(t)$ and 
$f_{SD}(t)=t^H$, which become straight lines in the linear-log 
representation 
of this figure.  The global air  temperature anomalies are 
characterized by a 
pdf scaling coefficient $\delta_a=0.87\pm0.02$ and a standard deviation 
scaling coefficient $H_a=0.92\pm0.01$. The global sea  temperature 
anomalies 
are characterized by a pdf scaling coefficient $\delta_s=0.89\pm0.02$ 
and a 
standard deviation scaling coefficient $H_s=0.94\pm0.01$.}

\end{figure}

\begin{table}
  \begin{tabular}{|c|c|c|c|}

               & $H$    & $\delta$ & $\mu$  \\ \hline   
Air         & $0.92\pm0.01 $       & $0.87\pm0.02$& $ 2.14\pm0.02$   \\ 
\hline
Sea         &  $0.94\pm0.01 $    & $0.89\pm0.02$ & $2.12\pm0.02 $ 

  \end{tabular} 

\caption{Exponents $H$, $\delta$ and $\mu$ for the global air  and sea   
temperature anomalies.}

\end{table}

\newpage

\begin{figure}

Figure 7\\
\epsfig{file=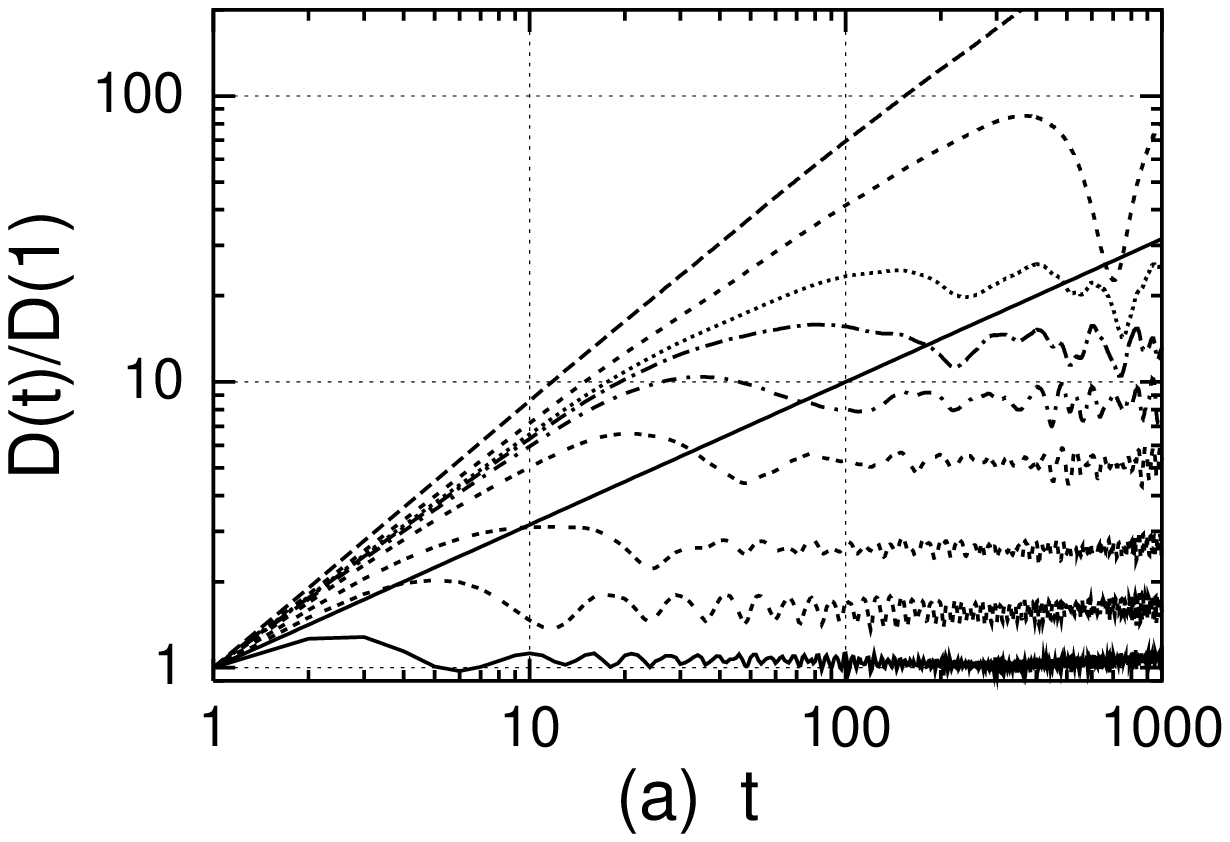,height=14cm,width=10cm,angle=-90}\\
\epsfig{file=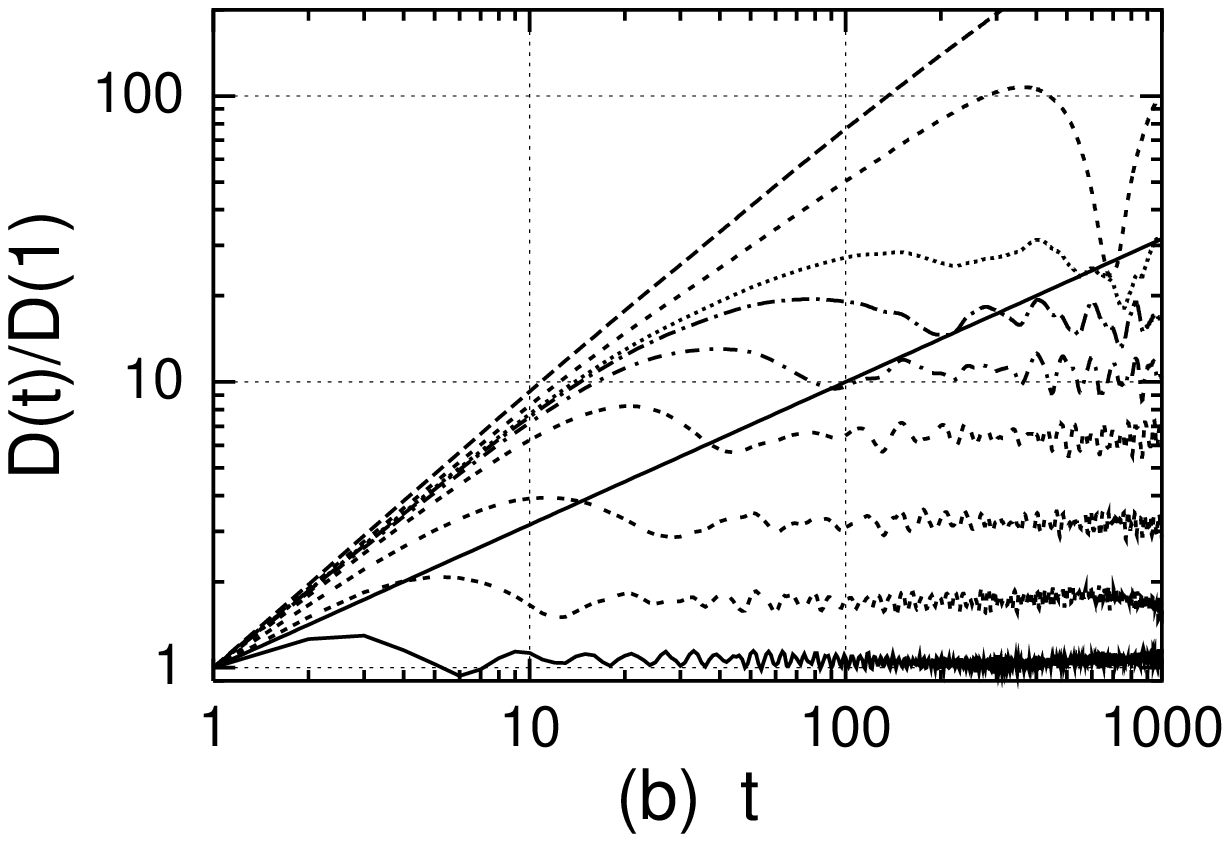,height=14cm,width=10cm,angle=-90}

\caption{SDA of the global air (a) and sea (b)  temperature anomalies 
of the 
residuals $R_j$.  From top to down, the curves are the SDA of (1) the 
original data, (2) R9, (3) R8,..., (9) R2. The straight line is 
$f_{SD}(t)=t^{0.5}$ that corresponds to the Gaussian diffusion.}

\end{figure}

\newpage

\begin{figure}

Figure 8\\
\epsfig{file=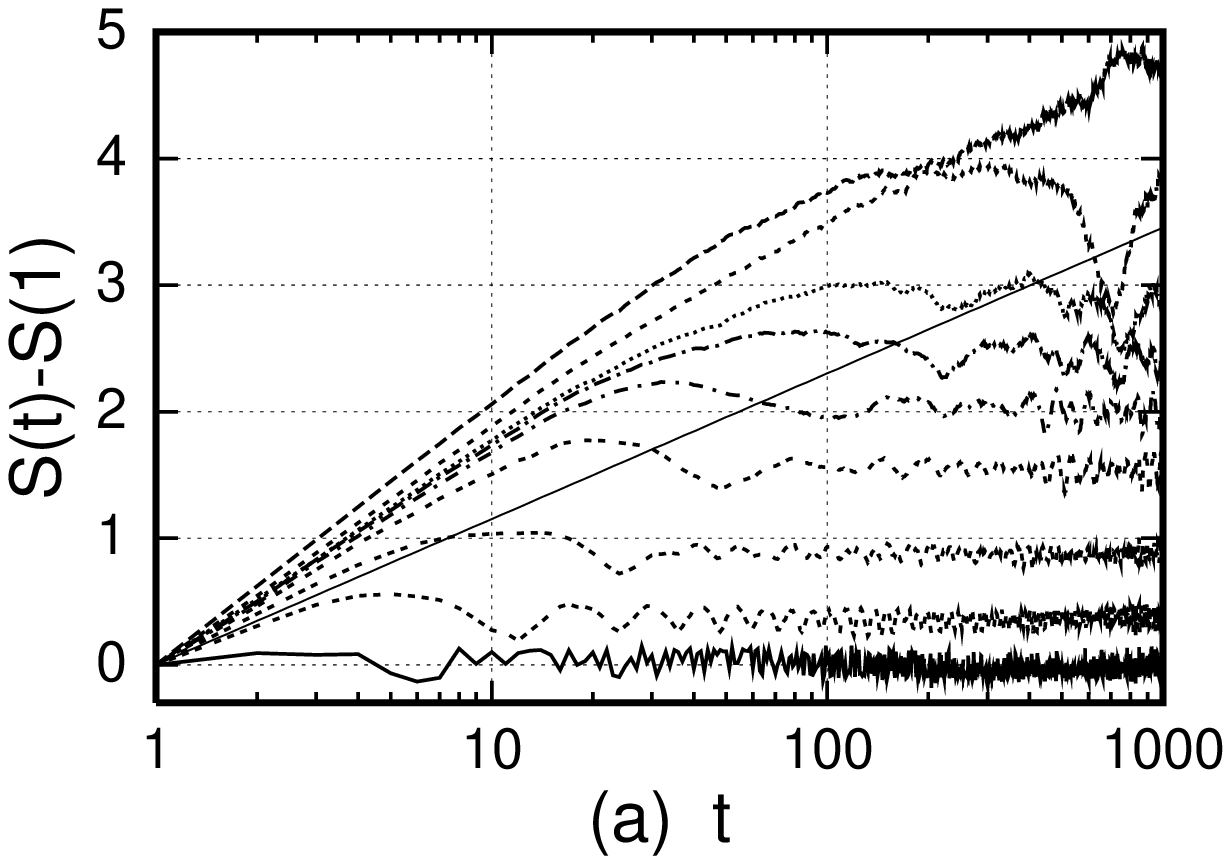,height=14cm,width=10cm,angle=-90}\\
\epsfig{file=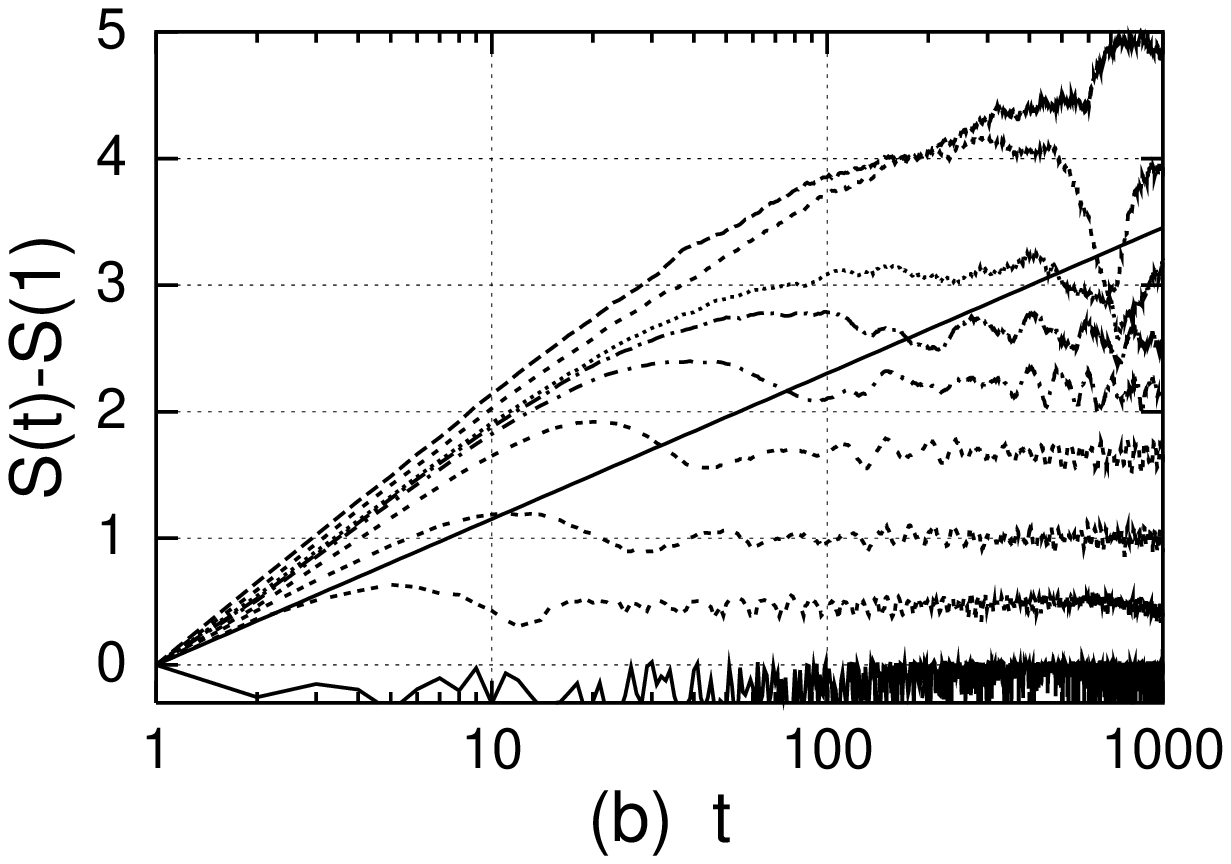,height=14cm,width=10cm,angle=-90}

\caption{DEA of the global air (a) and sea (b)  temperature anomalies 
of the 
residuals $R_j$.  From top to down, the curves are the SDA of (1) the 
original data, (2) R9, (3) R8,..., (9) R2. The straight line is 
$f_{DE}(t)={0.5}\ln(t)$  that corresponds to the Gaussian diffusion.}

\end{figure}

\newpage
\begin{table}
  \begin{tabular}{|c|c|c|c|c|c|c|c|c|c|c|}
    &D(1)   &D(1)   &S(1)   &S(1)   
&D$_{line}$&D$_{line}$&S$_{line}$&S$_{line}$&D-period&S-period      \\ 
    &Air    &Sea    &Air    &Sea    &Air    &Sea    &Air    &Sea    
&Air    
&Sea            \\ \hline
data    &0.243  &0.221  &1.44   &1.42   &   &   &   &   &   &       \\ 
\hline
R9  &0.162  &0.128  &1.09   &0.98   &23 &24 &2.80   &2.95   &707    
&686        
\\ \hline
R8  &0.158  &0.124  &1.07   &0.96   &14 &18 &2.50   &2.60   &238    
&221        
\\\hline    
R7  &0.152  &0.117  &1.03   &0.91   &11 &14 &2.27   &2.40   &220    
&211        
\\ \hline
R6  &0.140  &0.103  &0.96   &0.83   &7.5    &9.5    &1.95   &2.05   &96 
&93     
\\ \hline
R5  &0.124  &0.082  &0.85   &0.61   &4.5    &5.7    &1.40   &1.55   &48 
&48     
\\ \hline
R4  &0.112  &0.069  &0.82   &0.44   &2.35   &2.90   &0.78   &0.90   &24 
&24     
\\ \hline
R3  &0.092  &0.056  &0.69   &0.68   &1.45   &1.55   &0.25   &0.38   &12 
&12     
\\ \hline
R2  &0.066  &0.040  &0.43   &0.50   &   &   &   &   &6  &6  

  \end{tabular} 

\caption{ Summary of the information contained in  Figs. 7 and 8. The 
first 
four columns report the values of the standard deviation and of the 
entropy 
at the first step of diffusion for both global air and sea temperature 
anomalies and their rests $R_j$ for $j=2,...,9$. The four following 
columns, 
from the  5th to the 8th,  report the values of the height of the 
horizontal 
lines that measures the maximum  spreading (in the case of SDA, Figs. 
7) and 
the information or entropy (in the case of DEA, Figs. 8) that 
corresponds to 
each wavelet scale. The reported heights are relative to the values of 
the 
SDA and DEA at $l=1$. This means that the SDA and DEA heights  are 
defined by 
$D_{line}=D_j/D(1)$ and $S_{line}=S_j-S(1)$, respectively, where $S_j$ 
and 
$D_j$ are the values of the entropy and of the standard deviation of 
the 
diffusion process generated by the residuals $R_j$.  The last two 
columns 
report the main periodicities present in each residuals $R_j$. 
}

\end{table}

\newpage

\begin{figure}

Figure 9\\
\epsfig{file=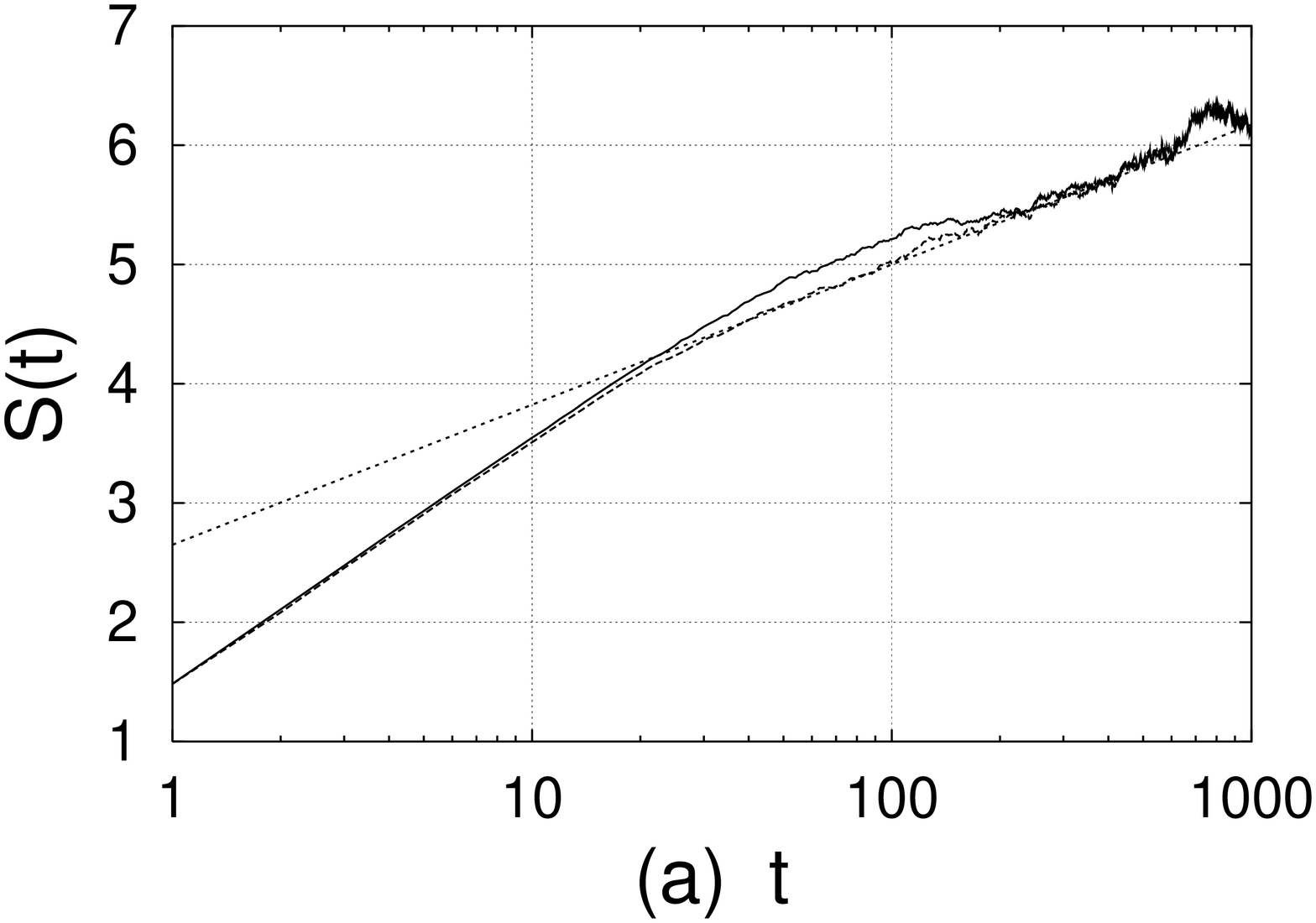,height=8.9cm,width=6cm,angle=-90}
\epsfig{file=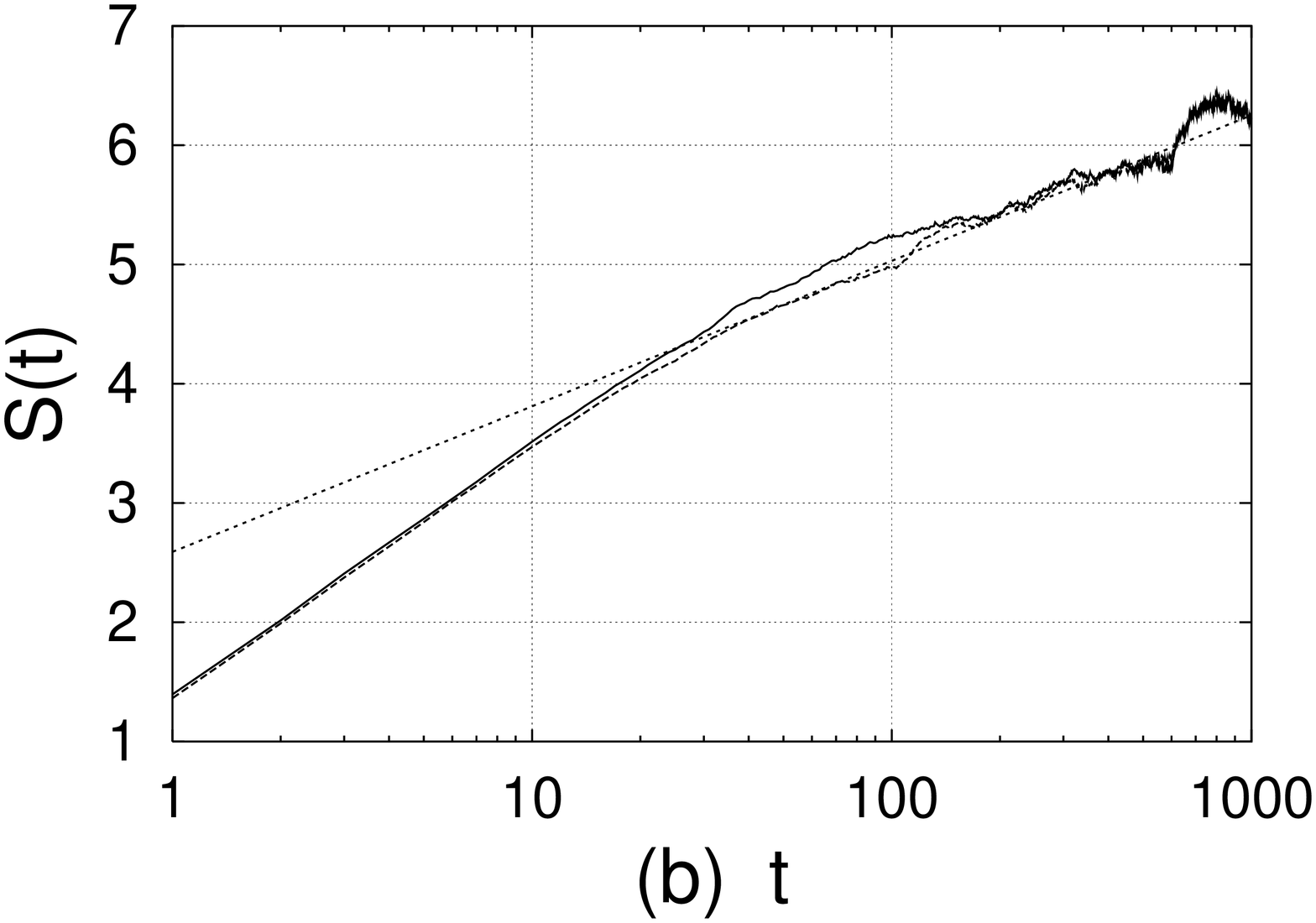,height=8.9cm,width=6cm,angle=-90} \\
\epsfig{file=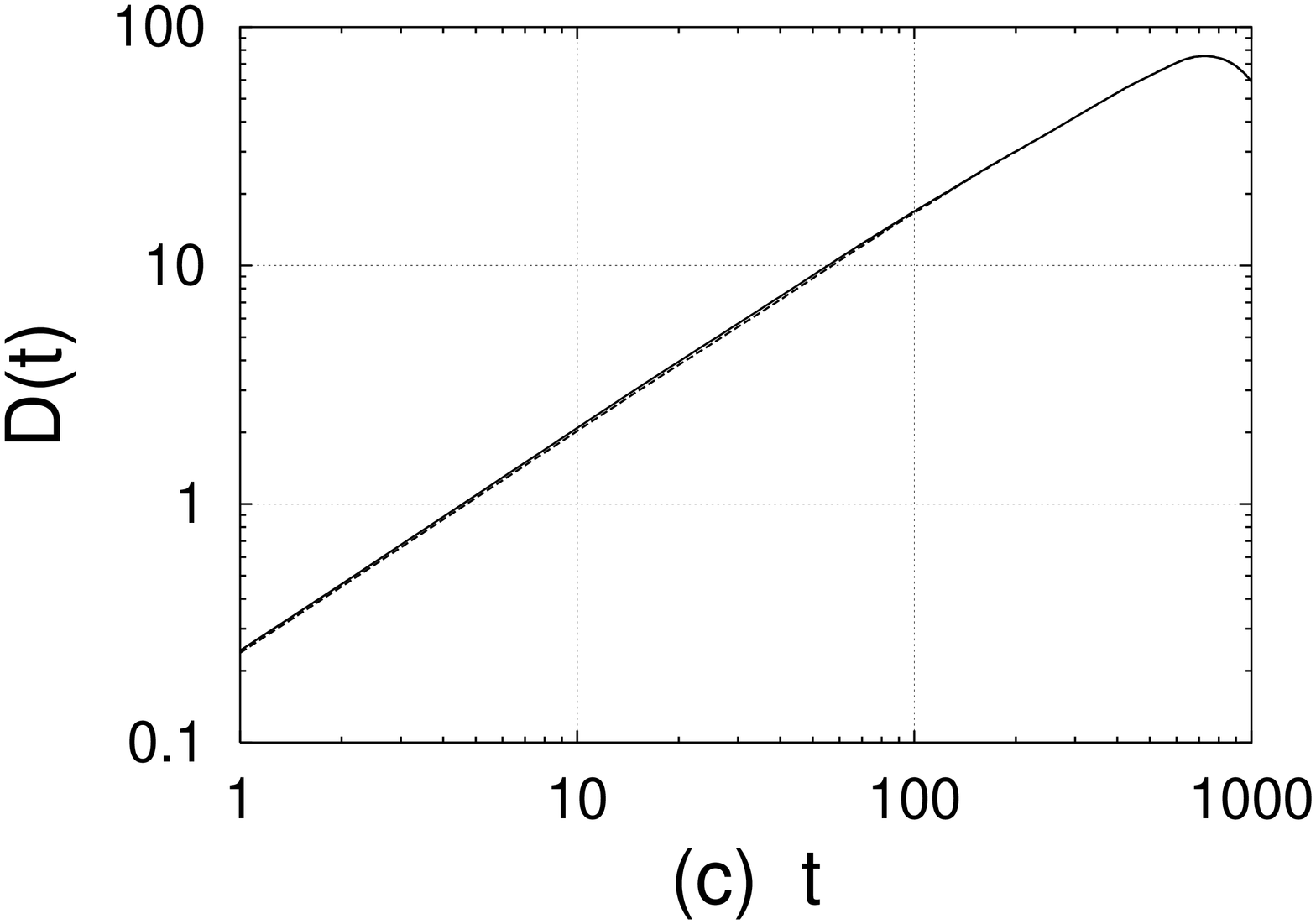,height=8.9cm,width=6cm,angle=-90}
\epsfig{file=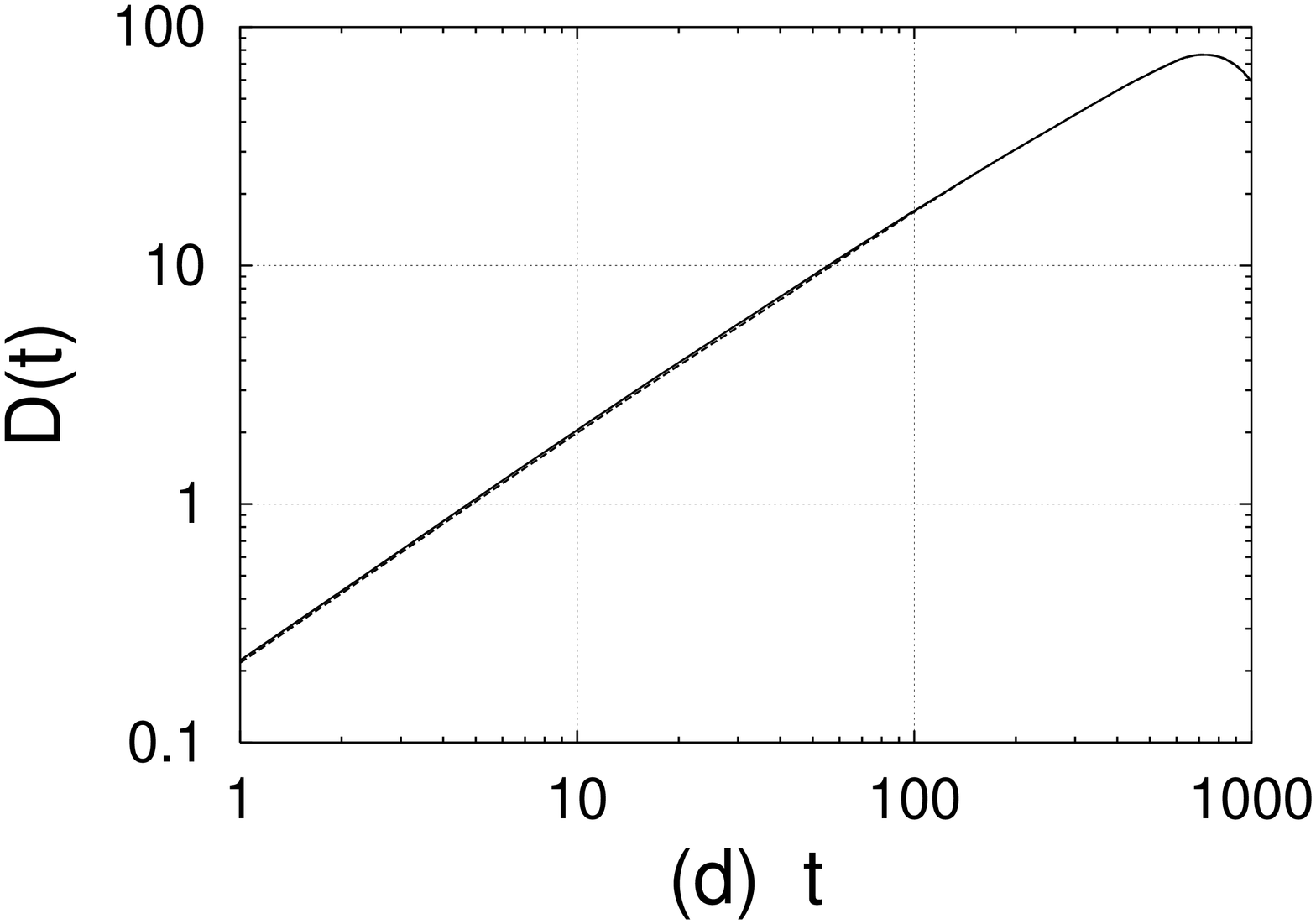,height=8.9cm,width=6cm,angle=-90}

\caption{DEA and SDA of the global air (a, c) and sea (b, d)  
temperature 
anomalies of the original data (solid line) and of the data detrended 
of the 
detail D7 (dashed line). The straight lines in (a) e (b) have a slope 
of 
$\delta=0.51\pm0.02$ (a) and $\delta=0.53\pm0.02$ (b) that correspond 
to the 
Gaussian diffusion. The effect of the detail D7 is detected by DEA but 
not by 
SDA.
}

\end{figure}

%%%%%%%%%%%%%%%%%%%%%%%%%%%%%%%%%%%%%%%%%%%%%%

   \end{document}